\documentclass[%twocolumn,%showpacs,preprintnumbers,
amsmath,%amssymb,aps,
prd,nofootinbib,floatfix,11pt,%preprint,
]{revtex4}

\usepackage{hyperref}
\usepackage{placeins}
\usepackage{afterpage}

\usepackage{graphicx}% Include figure files
\usepackage{setspace}
\usepackage{bm}% bold math

\usepackage[dvips]{color}
\newcommand\beq{\begin{eqnarray}}
\newcommand\eeq{\end{eqnarray}}

\def\lsim{\mathrel{\rlap{\lower4pt\hbox{$\sim$}}
    \raise1pt\hbox{$<$}}}                % less than or approx. symbol
\def\gsim{\mathrel{\rlap{\lower4pt\hbox{$\sim$}}
    \raise1pt\hbox{$>$}}}            

\allowdisplaybreaks
\newcommand\lnbar{\overline{\ln}}
\newcommand\MSbar{$\overline{\rm{MS}}$ }

\begin{document}

\renewcommand{\theequation}{\arabic{section}.\arabic{equation}}
\renewcommand{\thefigure}{\arabic{section}.\arabic{figure}}
\renewcommand{\thetable}{\arabic{section}.\arabic{table}}

\title{\Large \baselineskip=20pt 
Bare effective potential and Goldstone boson anti-resummation}
\author{Stephen P.~Martin}
\affiliation{\mbox{\it Department of Physics, Northern Illinois University, DeKalb IL 60115}}

\begin{abstract}\normalsize \baselineskip=16pt
I discuss the evaluation of the effective potential and related quantities starting from bare perturbation theory followed by \MSbar renormalization. The advantages of this method include a simple new way of avoiding infrared problems associated with Goldstone bosons, by treating their bare squared masses as two-point interaction vertices instead of incorporating them into the propagators. Thus, only positive integer powers of the Goldstone-boson squared masses appear at every stage of calculation, avoiding the imaginary parts and singularities encountered in intermediate steps of previous resummation methods for the minimization conditions. This Goldstone boson anti-resummation approach is implemented using simple rules provided through three-loop order in terms of the master vacuum integrals. Explicit results are given for the Standard Model at three-loop order. The method is shown to give effective potential minimization conditions that are consistent with previous resummation methods, but in a form more directly useful for related future multi-loop calculations of observables in the tadpole-free scheme.      
\end{abstract}

\maketitle

\tableofcontents

\baselineskip=15.7pt

%\newpage

%%%%%%%%%%%%%%%%%%%%%%%%%%%%%%%%%%%%%%%%%%%%%%%%%%%%%%%%%%%%%%%
\section{Introduction\label{sec:intro}}
\setcounter{equation}{0}
\setcounter{figure}{0}
\setcounter{table}{0} 
\setcounter{footnote}{1}

The effective potential \cite{Coleman:1973jx,Jackiw:1974cv,Sher:1988mj,Quiros:1999jp,Masina:2025pnp} in quantum field theory has many practical applications. It is an important tool in quantitatively understanding spontaneous symmetry breaking and vacuum selection, metastability and tunneling rates for false vacuum decay, phase transitions, associated types of gravitational wave production, baryogenesis, and inflation and evolution of the early universe. For general renormalizable theories, including supersymmetric ones, the zero-temperature effective potential has been given in Landau gauge at two-loop \cite{Ford:1992pn,Martin:2001vx} and three-loop \cite{Martin:2017lqn,Martin:2023fno} orders, and at two-loop order in a large class of more general gauge-fixing schemes in \cite{Martin:2018emo}. Landau gauge is often preferred for its simplicity, as it avoids mixing between scalar and vector propagators.

The motivations for the present paper stem from the pursuit of a program of precise calculations specifically within the Standard Model, which require an accurate relation between Lagrangian and observable quantities and the electroweak symmetry breaking Higgs boson vacuum expectation value (VEV). The effective potential for the Standard Model was first obtained at complete two-loop order in a groundbreaking paper by Ford, Jack, and Jones in \cite{Ford:1992pn} and extended to three loops in \cite{Martin:2013gka,Martin:2017lqn}, and to leading order in QCD at four loops in \cite{Martin:2015eia}.

In the Standard Model, the tree-level part of the renormalized effective potential for the real neutral component $\phi$ of the Higgs field is conventionally written as
\beq
V^{(0)} &=& \Lambda + \frac{1}{2} m^2 \phi^2 + \frac{1}{4} \lambda \phi^4,
\eeq
where $\Lambda$ is the field-independent vacuum energy (necessary for renormalization group invariance), $m^2$ is a negative squared-mass parameter, and $\lambda$ is the Higgs self-interaction coupling. The effective potential can now be obtained using dimensional regularization \cite{Bollini:1972ui,Bollini:1972bi,Ashmore:1972uj,Cicuta:1972jf,tHooft:1972tcz,tHooft:1973mfk}
in
\beq
d = 4 - 2\epsilon
\eeq
dimensions as a sum over 1-particle-irreducible loop diagrams in the presence of $\phi$-dependent masses and couplings. (For softly broken supersymmetric theories, dimensional reduction 
is used instead \cite{Siegel:1979wq,Capper:1979ns,Jack:1997sr,Siegel:1980qs,Avdeev:1982xy,Jack:1994rk,Stockinger:2005gx}.) 
After renormalization in the modified minimal subtraction ($\overline{\rm MS}$) \cite{Bardeen:1978yd,Braaten:1981dv} scheme, one writes the effective potential in a loop expansion as
\beq
V_{\rm eff}(\phi) = V^{(0)} + \kappa V^{(1)} + \kappa^2 V^{(2)} + \kappa^3 V^{(3)} + \cdots,
\label{eq:Veffphi}
\eeq
where the power of the quantity
\beq
\kappa = \frac{1}{16 \pi^2}
\eeq
is used in this paper to count the order of perturbation theory, coinciding with the number of loops.

The standard perturbative expansion in Landau gauge is plagued by infrared (IR) issues from Feynman diagrams involving the charged and neutral Goldstone boson components of the Higgs field, with a common tree-level squared mass denoted by 
\beq
G &=& m^2 + \lambda \phi^2
.
\eeq
Note that $G$ can also be written as 
\beq
G &=& \frac{1}{\phi} \frac{\partial V^{(0)}}{\partial \phi},
\label{eq:GfromV0}
\eeq
showing that while it does not vanish for general $\phi$, and in particular not at the minimum of the full $V_{\rm eff}(\phi)$ where we do calculations of other quantities, it would vanish if evaluated at the minimum of the tree-level potential. It is thus useful to keep in mind that $G$ can be thought of as effectively a loop-suppressed quantity,
%%\color{blue}%%NEW
at least near the minimum of the potential,
%%\color{black}%%ENDNEW
even though it appears in the tree-level Lagrangian.

Now, explicit calculation in the usual organization of perturbation theory shows that the Goldstone boson contributions to the effective potential have a leading behavior in $G$ like
\beq
V^{(1)} &\sim& G^2\hspace{1pt} \lnbar (G),
\\
V^{(2)} &\sim& G \hspace{1pt} \lnbar (G),
\\
V^{(3)} &\sim& \lnbar(G),
\\
V^{(n)} &\sim& \frac{1}{G^{n-3}}, \qquad \mbox{(for $n > 3$)} ,
\eeq
with
\beq
\lnbar(x) = \ln(x/Q^2),
\eeq
where $Q$ is the \MSbar renormalization scale.
After taking a derivative with respect to $\phi$ to find the minimum of $V_{\rm eff}$, one encounters logarithmically singular IR behavior as $G \rightarrow 0$ at two-loop order, with the problem apparently getting progressively worse, including inverse powers of $G$ at higher loop order. Furthermore, $G$ is negative in the Standard Model for reasonable choices of $Q$, so that the presence of $\lnbar(G)$ at each order in perturbation theory leads to an imaginary part of the effective potential. 
%%\color{blue}%%NEW
As was pointed out in ref.~\cite{Weinberg:1987vp}, the perturbatively calculated effective potential can have an imaginary part corresponding to a physical instability, but the full effective potential at its minimum should not have an imaginary part unless there is a genuine instability of the vacuum state. This is not the case in the Standard Model (although there could be a different sort of instability if the self-coupling runs negative in the far UV).
%%\color{black}%%ENDNEW
 This ``Goldstone boson catastrophe" is not confined to just the effective potential, as it arises in any multi-loop calculation involving Goldstone bosons. 

Resummation of the Goldstone boson contributions tames the problem, eliminating both $\lnbar(G)$ and negative powers of $G$, as shown in refs.~\cite{Martin:2014bca,Elias-Miro:2014pca}. The problem of gauge dependence was addressed in refs.~\cite{Andreassen:2014eha,Espinosa:2016uaw}, and 
further valuable elucidations have been made in refs.~\cite{Pilaftsis:2015bbs,Kumar:2016ltb,Braathen:2016cqe,Pilaftsis:2017enx,Braathen:2017izn,Espinosa:2017aew,Ekstedt:2018ftj}. 
One important practical outcome of these results involves the minimization condition for the full effective potential up to any fixed loop order $\ell$. From eqs.~(\ref{eq:Veffphi}) and (\ref{eq:GfromV0}), the minimization condition can be written as
\beq
G &=& -\sum_{n=1}^\ell \kappa^n \frac{1}{\phi} \frac{\partial V^{(n)}}{\partial \phi}
,
\eeq
where the terms on the right side depend on $G$, including the aforementioned logarithms for all $n$ and negative powers for $n\geq 3$. Resummation accomplishes a reorganization (up to neglected terms of order $\ell+1$) of this into the form 
\beq
G &=&  -\sum_{n=1}^\ell \kappa^n \Delta^{(n)},
\label{eq:GminDelta}
\eeq
where the $\Delta^{(n)}$ do not depend on $G$ at all. The
quantities $\Delta^{(n)}$ have been obtained up to three-loop order for the Standard Model in ref.~\cite{Martin:2017lqn}. Requiring that eq.~(\ref{eq:GminDelta}) holds corresponds to choosing a prescription for the VEV that one is expanding around, such that $V_{\rm eff}$ is minimized and the sum of all tadpoles vanishes up to loop order $\ell$. However, in the present paper I propose an improvement based on a better organization of perturbation theory.

The tadpole-free \MSbar scheme has been used in a continuing program of calculations of multi-loop accuracy for observables in the Standard Model in terms of the Lagrangian parameters, with results incorporated in the code library {\tt SMDR} \cite{Martin:2019lqd}, with a simplified and faster interpolation set of formulas in ref.~\cite{Alam:2022cdv}. These will be updated as new calculations become available. So far, the state-of-the-art results include the  complete two-loop and leading three-loop results for the pole masses of the Higgs, $W$, and $Z$ bosons \cite{Martin:2014cxa,Martin:2015lxa,Martin:2015rea,Martin:2022qiv}, and the top-quark pole mass through four-loop order in QCD \cite{Tarrach:1980up,Gray:1990yh,Melnikov:2000qh,Marquard:2015qpa}  with complete two-loop order corrections from \cite{Martin:2016xsp}. Also included in {\tt SMDR} are the effects of the Standard Model running of the gauge couplings at two-loop 
\cite{Jones:1974mm,Caswell:1974gg,Jones:1981we,Machacek:1983tz,Machacek:1983fi,Machacek:1984zw,Ford:1992pn,Luo:2002ey},
three-loop 
\cite{Tarasov:1980au,Larin:1993tp,Mihaila:2012fm,Mihaila:2012pz,Bednyakov:2012rb},
and four-loop
\cite{vanRitbergen:1997va,Czakon:2004bu,Bednyakov:2015ooa,Bednyakov:2016uia,Zoller:2015tha,Poole:2019txl,Davies:2019onf}
orders, and the strong coupling through five-loop order \cite{Baikov:2016tgj,Herzog:2017ohr,Luthe:2017ttg}, the running of Yukawa couplings at complete three-loop order \cite{Chetyrkin:2012rz,Bednyakov:2012en,Bednyakov:2014pia} with QCD corrections at four-loop  \cite{Chetyrkin:1997dh,Vermaseren:1997fq} and five-loop \cite{Baikov:2014qja} orders, and the 
running of $m^2$, $\lambda$, and $\phi^2$ through complete three-loop order \cite{Chetyrkin:2012rz,Chetyrkin:2013wya,Bednyakov:2013eba}, with four-loop corrections for the Higgs self coupling $\lambda$ at leading order in QCD \cite{Martin:2015eia,Chetyrkin:2016ruf}.
Also included are the matching rules \cite{Martin:2018yow} for decoupling of dimensionless couplings
at mass thresholds. The relation between the Lagrangian parameters and the Fermi decay constant is given at full three-loop order in \cite{Martin:2025cas}.

For context in the following, it is useful to very briefly describe the most efficient method for calculations in this program, which in the future will hopefully fill in the remaining three-loop contributions for the electroweak boson ($h$, $W$, and $Z$) pole masses. First, it is convenient to start by computing each observable in dimensional regularization in terms of bare parameters. This means that there are no counterterm diagrams. Instead, renormalization of the calculated quantity is simply carried out by a parameter redefinition.\footnote{If, as a student, you were required to calculate counterterm Feynman diagrams, without also being taught the simpler parameter-redefinition method of renormalization, then you should seek legal advice; you may be entitled to compensation.} 
Specifically, one rewrites the calculated formula for the quantity of interest by expressing each of the bare Lagrangian parameters (including VEVs) $X_B$ in terms of the corresponding renormalized parameters $X$. In the \MSbar scheme, the bare/renormalized relationships of Lagrangian quantities $X$ have the form
\beq
X_B &=& \mu^{\epsilon\rho^{\phantom{.}}_X } \left (X + \sum_{n=1}^\infty \kappa^n \sum_{k=1}^n 
\frac{c_{n,k}^X}{\epsilon^k} \right ),
\label{eq:XBrenorm}
\eeq
where the dimensional regularization mass scale $\mu$ introduced with each loop integration 
$\mu^{2 \epsilon} \int d^{4 - 2 \epsilon} p$ is related to the \MSbar renormalization scale $Q$ by
\beq
Q^2 &=& 4\pi e^{-\gamma_E} \mu^2,
\eeq
where $\gamma_E = 0.5772156649\ldots$ is Euler's constant. The counterterm coefficients $c_{n,k}^X$ are polynomials in the \MSbar renormalized parameters. In the Standard Model, the list of \MSbar parameters $X$ includes the gauge couplings $g_3$, $g$, and $g'$, the Higgs self coupling $\lambda$, the negative squared-mass parameter $m^2$, the field-independent vacuum energy $\Lambda$, the Yukawa couplings $y_t, y_b, \ldots$, and the Higgs background field $\phi$, with 
$\rho_\Lambda = -2$, and
$\rho_{g_3} = \rho_{g} = \rho_{g'} = \rho_{y_t} = \rho_{y_b} = \rho_{y_\tau} = 1$, 
and $\rho_\lambda = 2$, and $\rho_{m^2} = 0$, and $\rho_\phi = -1$, so that $\rho_G = 0$. 
Of course, the advantage of not needing to calculate any counterterm Feynman diagrams relies on the coefficients $c_{n,k}^X$ being known. But this is not an impediment, since they are independent of the quantity being calculated, and they follow immediately from the known beta functions of the theory
\beq
\beta_X &\equiv& Q \frac{d X}{dQ}\biggl |_{\epsilon = 0} 
\>=\> Q \frac{d X}{dQ} + \epsilon \rho_X X 
\>=\> \sum_{n=1}^\infty \kappa^n \beta_X^{(n)},
\eeq
through the identities
\beq
2 n c^X_{n,1} &=& \beta_X^{(n)},
\\
2 n c^X_{n,k} &=& \sum_{j=1}^{n-k+1} \sum_Y \beta_Y^{(j)} \frac{\partial}{\partial Y} c^X_{n-j,k-1}\qquad\mbox{(for $k>1$)}.
\eeq 
For convenience, the complete set of counterterm coefficients $c_{n,k}^X$ for the Standard Model through three-loop order are collected in the ancillary file {\tt SMcounters.txt} distributed with this paper. After collecting like powers of $\kappa$, the loop integrals in the calculated quantity of interest are then expanded in powers of $\epsilon$. The final step is to take the limit $\epsilon \rightarrow 0$ for the quantity of interest. For an IR-safe observable, the poles in $\epsilon$ must cancel as an essential consequence of renormalizability, providing useful consistency checks. Alternatively, the cancellation of poles for appropriate observables allows one to infer the coefficients $c_{n,k}^X$ and thus check the beta functions.

In the above-described method, there is also no need to calculate any tadpole diagrams,  because the sum of all tadpole graphs vanishes identically provided that one imposes the minimization condition for the full effective potential.\footnote{Another popular organization of perturbation theory instead expands around the VEV that minimizes the tree-level potential. This has the advantage that it allows maintaining manifest gauge invariance of the VEV and does not restrict to Landau gauge. However, it has two related major disadvantages. First, one must include tadpole subdiagrams explicitly in every calculation,
since the sum of them all does not vanish. Second,
perturbation theory converges more slowly. For example, in the Standard Model, the loop expansion parameter
associated with top-quark loops is not $N_c y_t^2/16 \pi^2$, but $N_c y_t^4/16 \pi^2 \lambda$, featuring an extra factor of $y_t^2/\lambda$. The Higgs 
self-coupling $\lambda$ denominators are due to tadpoles involving Higgs propagators at zero momentum. This parametrically slower convergence of perturbation theory is the penalty to be paid for expanding about the wrong VEV.}
Now, the absence of tadpole diagrams is realized most directly in the bare theory, so in the most straightforward approach it is the bare effective potential that should be minimized. To do so, one calculates the effective potential for a general renormalizable theory in the form
\beq
V_{\rm eff} = V_B^{(0)} + \kappa V_B^{(1)} + \kappa^2 V_B^{(2)} + \kappa^3 V_B^{(3)} + \cdots
,
\eeq
where the $V_B^{(n)}$ are written in terms of the bare Lagrangian parameters and unsubtracted loop integral functions in $d=4 - 2 \epsilon$ dimensions. In contrast, the corresponding results of refs.~\cite{Martin:2001vx,Martin:2017lqn} are organized in terms of renormalized parameters and renormalized loop integrals (in which subdivergences have been subtracted and the limit $\epsilon \rightarrow 0$ taken).

Consistent with this general approach is an improved method of dealing with the Goldstone boson problem. To illustrate the idea, consider the Standard Model with bare tree-level potential 
\beq
V^{(0)}_B &=&  \Lambda_B + \frac{1}{2} m_B^2 \phi_B^2 + \frac{1}{4} \lambda_B \phi_B^4,
\label{eq:Vbare0SM}
\eeq
and Goldstone-boson squared mass
\beq
G_B \,=\, m_B^2 + \lambda_B \phi_B^2 \,=\, \frac{1}{\phi_B} \frac{\partial V_B^{(0)}}{\partial \phi_B}.
\eeq
It follows that the minimization condition for $V_{\rm eff}$ in terms of the bare parameters (in order that the sum of all tadpoles vanishes identically) is
\beq
G_B &=& -\sum_{n=1}^\infty \kappa^n \frac{1}{\phi_B} \frac{\partial V_B^{(n)}}{\partial \phi_B}
.
\label{eq:GBdeltaBnOLD}
\eeq
Note that in $4 - 2 \epsilon$ dimensions, with the standard organization of perturbation theory, 
the terms on the right side of eq.~(\ref{eq:GBdeltaBnOLD}) contain fractional powers of $G_B$, for example 
$G_B^{3 - n - \epsilon}$ from circular ``daisy-chain'' diagrams at $n$-loop order containing $n-1$ Goldstone propagators all carrying the same momentum, separated by $n-1$ one-loop sub-diagrams containing heavy particles. This is the realization of the Goldstone boson catastrophe in the bare perturbation theory, which we want to improve.

In this paper, I propose an embarrassingly simple solution: all Goldstone-boson squared mass terms in the bare Lagrangian are to be treated as two-point interaction vertices, rather than being incorporated in the propagators. This amounts to an organization of perturbation theory such that, compared to the usual treatment, one makes the replacement
\beq
\frac{1}{p^2 + G_B} &\rightarrow& \frac{1}{p^2} - \frac{G_B}{(p^2)^2} + \frac{G_B^2}{(p^2)^3} - \frac{G_B^3}{(p^2)^4} + \cdots
\label{eq:expandGBprop}
\eeq
for every internal Goldstone boson with 4-momentum $p^\mu$. (The metric signature is Euclidean or $-$$+$$+$$+$.)
In other words, all Goldstone bosons are treated as having exactly 0 bare mass and a two-particle interaction vertex proportional to $G_B$. One does {\em not} resum the infinite series in eq.~(\ref{eq:expandGBprop}), but instead associates each power of $G_B$ with an additional loop factor.
It bears emphasis that the arrow in eq.~(\ref{eq:expandGBprop})  does not represent a diagram-by-diagram equality, but rather a prescription for the transformation of one consistent organization of perturbation theory to another (better) one. I will return to this point in section \ref{sec:Goldstonerules}.

It follows immediately from eq.~(\ref{eq:expandGBprop}) that the effective potential, and all other quantities computed with this prescription, have power series expansions containing only non-negative integer powers of $G_B$. In particular, the $n$-loop contribution to the bare effective potential takes the form
\beq
V_B^{(n)} &=& \sum_{j=0}^\infty G_B^j  V_B^{(n,j)} .
\eeq
Now, using 
\beq
\frac{1}{\phi_B} \frac{\partial G_B}{\partial \phi_B} = 2 \lambda_B,
\eeq
the minimization condition eq.~(\ref{eq:GBdeltaBnOLD}) is replaced by
\beq
G_B &=& -\sum_{n=1}^\infty \kappa^n \delta_B^{(n)},
\label{eq:GBdeltaBn}
\eeq
where
\beq
\delta_B^{(n)} &=& 
\frac{1}{\phi_B} \frac{\partial V_B^{(n)}}{\partial \phi_B} 
\,=\, 
\sum_{j=0}^\infty G_B^j \delta_B^{(n,j)},
\eeq
with
\beq
\delta_B^{(n,j)} 
&=& 
\frac{1}{\phi_B} \frac{\partial V_B^{(n,j)}}{\partial \phi_B} 
+ 2 (j+1) \lambda_B V_B^{(n,j+1)} .
\label{eq:deltaBnj}
\eeq
The minimization condition eq.~(\ref{eq:GBdeltaBn}) therefore becomes the nonlinear equation
\beq
G_B &=& -\sum_{n=1}^\infty \sum_{j=0}^\infty \kappa^n \hspace{1pt} G^j_B\hspace{1pt} \delta^{(n,j)}_B .
\eeq
For this equation it is straightforward to obtain, by substitution followed by matching powers of $\kappa$, the solution
\beq
G_B &=&  -\sum_{n=1}^\infty \kappa^n \Delta_B^{(n)} ,
\label{eq:GBminDeltaB}
\eeq
where the quantities on the right side are independent of $G_B$ (or equivalently $m_B^2$). Through four-loop order they are given by 
\beq
\Delta_B^{(1)} &=& \delta_B^{(1,0)},
\label{eq:DeltaBonedeltaB}
\\
\Delta_B^{(2)} &=& \delta_B^{(2,0)} - \delta_B^{(1,0)} \delta_B^{(1,1)},
\label{eq:DeltaBtwodeltaB}
\\
\Delta_B^{(3)} &=& \delta_B^{(3,0)} - \delta_B^{(2,0)} \delta_B^{(1,1)}
- \delta_B^{(2,1)} \delta_B^{(1,0)}
+ \delta_B^{(1,2)} \left [ \delta_B^{(1,0)} \right ]^2
+ \delta_B^{(1,0)} \left [ \delta_B^{(1,1)} \right ]^2
,
\label{eq:DeltaBthreedeltaB}
\\
\Delta_B^{(4)} &=& \delta_B^{(4,0)}
- \delta_B^{(3,1)} \delta_B^{(1,0)}
- \delta_B^{(3,0)} \delta_B^{(1,1)}
- \delta_B^{(2,1)} \delta_B^{(2,0)}
+ \delta_B^{(2,2)} \left [ \delta_B^{(1,0)} \right ]^2
\nonumber \\ && 
+ \delta_B^{(2,0)} \left [ \delta_B^{(1,1)} \right ]^2
+ 2 \delta_B^{(2,1)} \delta_B^{(1,1)} \delta_B^{(1,0)}
+ 2 \delta_B^{(2,0)} \delta_B^{(1,2)} \delta_B^{(1,0)}
\nonumber \\ && 
- \delta_B^{(1,0)} \left [ \delta_B^{(1,1)} \right ]^3
- \delta_B^{(1,3)} \left [ \delta_B^{(1,0)} \right ]^3
- 3 \delta_B^{(1,2)} \delta_B^{(1,1)} \left [ \delta_B^{(1,0)} \right ]^2
.
\eeq
Equation (\ref{eq:GBminDeltaB}) is the condition that must be imposed between the bare VEV $\phi_B$ and the bare $m_B^2$ (and used to eliminate the latter in favor of the former) in order to have a tadpole-free bare perturbation theory. When one calculates some other quantity in tadpole-free bare perturbation theory, $G_B$ is replaced by the right side of eq.~(\ref{eq:GBminDeltaB}), through the appropriate loop order. If one is calculating a quantity in which the minimization condition to eliminate $G_B$ is needed at loop order $\ell$, these equations show that one must evaluate $\delta^{(n,j)}_B$, and therefore expand $V_B^{(n)}$ through order $G_B^{j+1}$, for all $1 \leq n \leq \ell$ and $j\leq \ell - n$. This expansion is straightforwardly (in principle) done by calculating Feynman diagrams using eq.~(\ref{eq:expandGBprop}).

Since \MSbar renormalization is just a parameter redefinition with only non-negative integer powers of the renormalized quantities (in fact, polynomial at a given fixed loop order), the resulting calculation of the renormalized effective potential and its derivatives is also manifestly completely free of all logarithms and negative powers of the renormalized $G$.
%%\color{blue}%%NEW
The renormalized $V_{\rm eff}$ does have poles in $\epsilon$ corresponding to IR divergences, as we will note explicitly at the end of section \ref{sec:StandardModel}. This is not a problem because the effective potential itself is not an observable. The important point is that these IR poles in $\epsilon$ do not infect expressions for observable quantities.
%%\color{black}%%ENDNEW

As a check, which will be described in section \ref{sec:StandardModel}, in the case of the Standard Model the renormalized tadpole-free condition (\ref{eq:GminDelta}) found in previous papers follows from renormalization of the corresponding bare tadpole-free condition eq.~(\ref{eq:GBminDeltaB}). This amounts to substituting the bare parameters in terms of the \MSbar quantities as in eq.~(\ref{eq:XBrenorm}), followed by taking the limit $\epsilon \rightarrow 0$. In other words, the renormalized minimization condition for $G$ is correctly recovered from the bare minimization condition for $G_B$. As a nontrivial part of this check, all poles in $\epsilon$ cancel in the minimization condition, including IR poles. 

Despite this agreement, there is an important difference between eqs.~(\ref{eq:GminDelta}) and (\ref{eq:GBminDeltaB}), which makes the latter more useful in practice. The point is that when incorporating the minimization condition in the calculation of some other observable, in bare perturbation
theory $G_B$ may be multiplied by bare quantities and/or by integrals that themselves have poles in $\epsilon$. Therefore, when working in bare perturbation theory, when substituting for $G_B$ it is necessary to keep some terms with positive powers of $\epsilon$ in $\Delta_B^{(n)}$, which are not incorporated in the renormalized quantities $\Delta^{(n)}$ due to having taken the limit $\epsilon \rightarrow 0$. In the two-loop calculations of refs.~\cite{Martin:2014cxa,Martin:2015lxa,Martin:2015rea}, this issue was finessed by recognizing the necessity of the simple extra terms (due to essentially the argument just given) and including them by hand. In a sense, the exposition given here is a simple principled justification for this, which becomes comparatively much more useful for calculations at three-loop order and beyond. 
 
To reiterate, the method proposed here does not involve resummation of series of diagrams or subsets of diagrams, but rather simply organizes perturbation theory in such a way as to treat the Goldstone-boson bare squared masses strictly as interaction vertices from the start. Rather than unifying contributions from infinite series of different Feynman diagrams as in the usual resummation methods, in the present approach each Feynman diagram in the usual perturbation theory is deconstructed into a series of Feynman diagrams with exactly massless bare propagators. To emphasize this difference from other resummation schemes, I refer to the method described here as {\em Goldstone boson anti-resummation}. 

Importantly, the method applies in the same straightforward way to other calculations, including pole masses and even scattering amplitudes and decay rates, with the same feature that there are no logarithms or fractional or negative powers of the bare $G_B$ or the renormalized $G$ at any stage. It will be used in forthcoming papers to calculate previously unknown, and quite complicated, three-loop contributions to the pole masses of the Standard Model electroweak gauge bosons $h$, $W$, and $Z$. Indeed, this partially completed experience was the instigation for the present paper. 

\section{Bare effective potential functions\label{sec:bare}}
\setcounter{equation}{0}
\setcounter{figure}{0}
\setcounter{table}{0} 
\setcounter{footnote}{1}

Consider the bare effective potential at two-loop order in Landau gauge, for a general renormalizable field theory. The notation here closely follows that of ref.~\cite{Martin:2017lqn}, but with one important difference:
here, the Lagrangian parameters are bare parameters, each carrying a subscript $B$ as in eq.~(\ref{eq:XBrenorm}). After diagonalization of the squared-mass matrices (if necessary, by a unitary rotation change of basis of the fields), the field content consists of 
real scalars $R_i$ carrying lowercase-letter indices $i,j,k,\ldots$ from the middle of the alphabet, two-component fermions $\psi_I$ with capital-letter indices $I,J,K,\ldots$ from the middle of the alphabet, and real vectors $A_\mu^a$ with lowercase-letter indices $a,b,c$ from the beginning of the alphabet, with corresponding ghosts $\omega^a$ and antighosts $\overline \omega^a$. After diagonalizing the squared-mass terms in the presence of the scalar background fields, the bare interaction Lagrangian has the form
\beq
{\cal L}_{\rm int} &=& 
-\frac{1}{6} \lambda_B^{ijk} R_i R_j R_k - \frac{1}{24} \lambda_B^{ijkl} R_i R_j R_k R_l 
- \frac{1}{2} \left ( Y_B^{iJK} R_i \psi_J \psi_K + {\rm c.c.} \right ) 
\nonumber \\ &&
+ \left (g^{aJ}_{I} \right )_B A^{\mu a} \psi^{\dagger I} \overline \sigma_\mu \psi_J
- g_B^{ajk} A^{\mu a} R_j \partial_\mu R_k 
- \frac{1}{4} g_B^{abjk} A_\mu^a A^{\mu b} R_j R_k 
- \frac{1}{2} g_B^{ab j} A^a_\mu A^{\mu b} R_j
\nonumber \\ &&
- g_B^{abc} A^{\mu a} A^{\nu b} \partial_\mu A^c_\nu
- \frac{1}{4} g_B^{abe} g_B^{cde} A^{\mu a} A^{\nu b} A^c_\mu A^d_\nu
- g_B^{abc} A^{\mu a} \omega^b \partial_\mu \overline \omega^c .
\eeq
Here the background-field-dependent couplings are bare real scalar interactions $\lambda_B^{ijk}$, $\lambda_B^{ijkl}$, complex
Yukawa couplings $Y_B^{iJK}$ and fermionic gauge couplings $(g^{a J}_{I})_B$, real vector gauge interactions $g_B^{abc}$, and vector-scalar interactions $g_B^{ajk}$
and $g_B^{abj}$, and
\beq
g_B^{abjk} = g_B^{ajl} g_B^{bkl} + g_B^{akl} g_B^{bjl} .
\eeq
The two-component fermions have bare mass matrices $M^{II'}_B$. These are not necessarily diagonal (notably, they are off-diagonal if the fermions carry conserved charges), but 
the basis of fermionic fields is chosen by a unitary transformation so that the bare fermion squared-mass matrix 
$\bigl (M_{II'} \bigr )_B M_B^{I'J} = (m^{2J}_{I})_B$ is diagonal. 
By convention, flipping the heights of fermion indices corresponds 
to complex conjugation, so that
$\bigl (Y_{jIJ} \bigr )_B \equiv \bigl (Y_B^{jIJ} \bigr )^*$,
and 
$\bigl (M_{II'} \bigr )_B \equiv \bigl (M^{II'}_B \bigr )^*$,
and
$\bigl (g^{aI}_J \bigr )_B \equiv \bigl (\bigl (g^{aJ}_I \bigr )_B\bigr )^*$.
Repeated indices are implicitly summed when they appear only on one side of an equation.

The two master integrals needed to write the bare effective potential through two-loop order are 
\beq
{\bf A}(x) &=& \int_p \frac{1}{p^2 + x} \,=\, \Gamma(-1 + \epsilon) x (4 \pi \mu^2/x)^\epsilon
,
\\
{\bf I}(x,y,z) &=& \int_p \int_q \frac{1}{[p^2 + x][q^2 + y][(p-q)^2 + z]},
\eeq
where the Euclidean momentum integrations are normalized as
\beq
\int_p &=& (16 \pi^2) \frac{\mu^{2 \epsilon}}{(2 \pi)^d}\int d^d p .
\eeq
The evaluation and expansion in $\epsilon$ of the function ${\bf I}(x,y,z)$ was given in \cite{Ford:1992pn}.

The one-loop effective potential in the bare scheme can be written as
\beq
V_B^{(1)} &=& \sum_i {\bf f}(i) - 2 \sum_I {\bf f}(I) + (d-1) \sum_a {\bf f}(a),
\eeq
where the indices $i$, $I$, and $a$ are used to denote the bare squared-mass eigenvalues of the corresponding fields when appearing as arguments of functions, and
\beq
{\bf f}(x) &=& x {\bf A}(x)/d .
\eeq
The two-loop bare effective potential can similarly be written as
\beq
V^{(2)}_B &=& 
\frac{1}{8} \lambda_B^{jjkk} \hspace{1pt} {\bf f}_{SS}(j,k)
+ \frac{1}{12} \bigl (\lambda_B^{jkl} \bigr )^2 \hspace{1pt} {\bf f}_{SSS}(j,k,l) 
+\frac{1}{2} Y_B^{jIJ} \bigl (Y_{jIJ} \bigr )_B \hspace{1pt} {\bf f}_{FFS}(I,J,j)
\nonumber \\ &&
+\frac{1}{4} \left [Y_B^{jIJ} Y_B^{jI'J'} \bigl (M_{II'} \bigr )_B \bigl (M_{JJ'} \bigr )_B + {\rm c.c.} \right ] 
\hspace{1pt} {\bf f}_{\overline F \overline F S}(I,J,j)
+\frac{1}{4} \bigl (g_B^{ajk} \bigr )^2  \hspace{1pt} {\bf f}_{VSS}(a,j,k)
\nonumber \\ &&
+\frac{1}{4} \bigl (g_B^{abj} \bigr )^2  \hspace{1pt} {\bf f}_{VVS}(a,b,j)
+\frac{1}{2} \bigl (g^{aJ}_I \bigr )_B \bigl (g^{aI}_J \bigr )_B \hspace{1pt} {\bf f}_{FFV}(I,J,a)  
\nonumber \\ &&
+\frac{1}{2} \bigl (g^{aJ}_I \bigr )_B \bigl (g^{aJ'}_{I'} \bigr )_B M_B^{II'} \bigl (M_{JJ'} \bigr )_B \hspace{1pt} {\bf f}_{\overline F \overline F  V}(I,J,a)
+ \frac{1}{12} \bigl (g_B^{abc} \bigr )^2 \hspace{1pt} {\bf f}_{{\rm gauge}}(a,b,c)  
,
\label{eq:V2loopgeneral}
\eeq
where  the loop integral functions are
\beq
{\bf f}_{SS}(x,y) &=& {\bf A}(x) {\bf A}(y),
\label{eq:fSSxy}
\\
{\bf f}_{SSS}(x,y,z) &=& -{\bf I}(x,y,z),
\label{eq:fSSSxyz}
\\
{\bf f}_{FFS}(x,y,z) &=&  (x + y - z) {\bf I}(x, y, z) 
+ {\bf A}(x) {\bf A}(y) 
- {\bf A}(x) {\bf A}(z) 
- {\bf A}(y) {\bf A}(z)
,
\\
{\bf f}_{\overline F \overline FS}(x,y,z) &=& 2 {\bf I}(x,y,z)
,
\\
{\bf f}_{VSS}(x,y,z) &=& \frac{1}{x} \Bigl [-\lambda(x,y,z) {\bf I}(x, y, z) + (y-z)^2 {\bf I}(0, y, z)
+ x {\bf A}(y) {\bf A}(z) 
\nonumber \\ && 
+ \bigl \{(d-2) x - y + z \bigr \} {\bf A}(x) {\bf A}(y)
+ \bigl \{(d-2) x + y - z \bigr \} {\bf A}(x) {\bf A}(z)
\Bigr ]
,
\label{eq:fVSSxyz}
\\
{\bf f}_{VVS}(x,y,z) &=& \frac{1}{4 x y} \Bigl [ \bigl \{4 (1-d) x y -\lambda(x,y,z)  \bigr \} {\bf I}(x,y,z)
-z^2 {\bf I}(0,0,z) 
\nonumber \\ && 
+ (x-z)^2 {\bf I}(0,x,z) + (y-z)^2 {\bf I}(0,y,z) + (z - x - y) {\bf A}(x) {\bf A}(y) 
\nonumber \\ && 
+ y {\bf A}(x) {\bf A}(z) + x {\bf A}(y) {\bf A}(z) \Bigr ]
,
\label{eq:fVVSxyz}
\\
{\bf f}_{FFV}(x,y,z) &=& \frac{1}{z} \Bigl [\bigl \{ (x-y)^2 + (d-3) (x+y) z + (2-d) z^2 \bigr \} {\bf I}(x,y,z)
- (x-y)^2 {\bf I}(0,x,y) 
\nonumber \\ && 
+ \bigl \{ x-y + (2-d) z \bigr \}{\bf A}(x) {\bf A}(z) 
+ \bigl \{ y-x + (2-d) z \bigr \}{\bf A}(y) {\bf A}(z) 
\nonumber \\ && 
+ (d-2) z {\bf A}(x) {\bf A}(y) 
\Bigr ]
,
\label{eq:fFFVMS}
\\
{\bf f}_{\overline F \overline F V}(x,y,z) &=& 2 (d-1) {\bf I}(x, y, z)
,
\\
{\bf f}_{\rm gauge} (x,y,z) &=&  \frac{1}{4xyz} \Bigl [
\lambda(x,y,z) \bigl \{ 4 (1-d) (x y + x z + y z) - \lambda(x,y,z)  \bigr \}  {\bf I}(x,y,z) 
\nonumber \\ && 
+ (x-y)^2 \bigl \{(x-y)^2 + 4 (d-1) x y \bigr \} {\bf I}(0,x,y) + z^2 (2 x y - z^2) {\bf I}(0,0,z)
\nonumber \\ && 
+ (x-z)^2 \bigl \{(x-z)^2 + 4 (d-1) x z \bigr \} {\bf I}(0,x,z) + y^2 (2 x z - y^2) {\bf I}(0,0,y)
\nonumber \\ && 
+ (y-z)^2 \bigl \{(y-z)^2 + 4 (d-1) y z \bigr \} {\bf I}(0,y,z) + x^2 (2 y z - x^2) {\bf I}(0,0,x)
\nonumber \\ && 
+ z \bigl \{ (7 - 4 d) (x+y) (x + y - z)  + 4 d (d-2) x y  + z^2 \bigr \} {\bf A}(x) {\bf A}(y)
\nonumber \\ && 
+ y \bigl \{ (7 - 4 d) (x+z) (x + z - y)  + 4 d (d-2) x z  + y^2 \bigr \} {\bf A}(x) {\bf A}(z)
\nonumber \\ && 
+ x \bigl \{ (7 - 4 d) (y+z) (y + z - x)  + 4 d (d-2) y z  + x^2 \bigr \} {\bf A}(y) {\bf A}(z)
\Bigr ]
.
\label{eq:fgaugexyz}
\eeq
Equations (\ref{eq:fVSSxyz}), (\ref{eq:fVVSxyz}), and (\ref{eq:fgaugexyz}) employ the triangle function defined by
\beq
\lambda(x,y,z) &=& x^2 + y^2 + z^2 - 2 x y - 2 x z - 2 y z.
\eeq
Equations (\ref{eq:fVSSxyz})-(\ref{eq:fgaugexyz}) appear to have singularities whenever a bare vector squared mass $x$, $y$, or $z$ vanishes, due to the Landau gauge propagator form
\beq
\frac{1}{i} \frac{\eta^{\mu\nu} - p^\mu p^\nu/p^2}{p^2 + x},
\eeq
which contains the partial fraction decomposition
\beq
\frac{1}{p^2 (p^2 + x)} = \frac{1}{x} \left ( \frac{1}{p^2} - \frac{1}{p^2 + x} \right ) .
\eeq
However, this does not give rise to any actual singularity in the effective potential at 2-loop order, because
there are no doubled massless propagators that could cause IR problems. Instead, the integrals for those special cases with massless vectors that are not immediately obvious are\footnote{Note the amusing fact that ${\bf f}_{FFV}(x,y,0) = 0$. This seems surprising, in the sense that I do not know any reason why it had to be true. Note that it is not true for regularization by dimensional reduction; see eq.~(\ref{eq:fFFVxy0DRED}) below.} 
\beq
{\bf f}_{VSS}(0,x,y) &=& (d-1) \Bigl [(x + y)  {\bf I}(0, x, y) + {\bf A}(x) {\bf A}(y) \Bigr ] 
,
\label{eq:fVSS0xy}
\\
{\bf f}_{VVS}(0,x,y) &=& \frac{d-1}{4x} \Bigl [ 
(y - 3 x) {\bf I}(0,x,y) 
- y {\bf I}(0,0,y) 
+ {\bf A}(x) {\bf A}(y) 
\Bigr ]
,
\\
{\bf f}_{VVS}(0,0,x) &=& \frac{1}{4} d (1-d) {\bf I}(0,0,x)
,
\label{eq:fVVS00x}
\\
{\bf f}_{FFV}(x,y,0) &=& 0
,
\\
{\bf f}_{\rm gauge} (0,x,y) &=&  
\frac{1}{4xy} \Bigl [
(x+y) \bigl \{ (5 - 3 d)(x - y)^2 + 4 (d-1)^2 x y  \bigr \} {\bf I}(0,x,y) 
\nonumber \\ && 
+ x^2 \bigl \{ 2 y + (3 d - 5) x \bigr \}  {\bf I}(0,0,x) 
+ y^2 \bigl \{ 2 x + (3 d - 5) y \bigr \}  {\bf I}(0,0,y)
\nonumber \\ && 
+ \bigl \{ (5 - 3 d) (x-y)^2 + 4 (d-3)(2d-3) x y \bigr \} {\bf A}(x) {\bf A}(y) 
\Bigr ]
,
\\
{\bf f}_{\rm gauge} (0,0,x) &=& \frac{1}{4} (16 - 19d + 7 d^2) x {\bf I}(0,0,x)
.
\label{eq:fgauge00x}
\eeq
For convenience, the functions appearing in eqs.~(\ref{eq:fSSxy})-(\ref{eq:fgaugexyz}) and (\ref{eq:fVSS0xy})-(\ref{eq:fgauge00x}) are also given in the ancillary file {\tt bare2loop.txt} distributed with this paper.

In applications to softly broken supersymmetry, one uses regularization by dimensional reduction, instead of dimensional regularization. The essential difference is that although momentum integrations are continued to
$d = 4 - 2 \epsilon$ dimensions, vector bosons still have exactly 4 components. This has the consequence that the functions above that involve vector bosons are modified. Following ref.~\cite{Martin:2001vx}, I use capital ${\bf F}$ to distinguish the dimensional reduction counterparts in softly broken supersymmetry from the dimensional regularization functions ${\bf f}$ given above. The results for the 2-loop functions are
\beq
{\bf F}_{FFV}(x,y,z) &=& \frac{1}{z} \Bigl [ 
\bigl \{ (x-y)^2 + (x+y) z -2 z^2 \bigr \}\hspace{1pt} {\bf I}(x,y,z)
- (x-y)^2\hspace{1pt} {\bf I}(0,x,y) 
\nonumber \\ && 
+ 2 z {\bf A}(x) {\bf A}(y) 
+ \bigl ( x-y - 2 z \bigr ) {\bf A}(x) {\bf A}(z) 
+ \bigl ( y-x - 2 z \bigr ) {\bf A}(y) {\bf A}(z) 
\Bigr ]
,
\label{eq:FFFVxyz}
\\
{\bf F}_{FFV}(x,y,0) &=& (4-d) \left [ {\bf A}(x) {\bf A}(y) + (x+y) {\bf I}(0, x, y) \right ],
\label{eq:fFFVxy0DRED}
\\
{\bf F}_{\overline F \overline FV}(x,y,z) &=& 6 {\bf I}(x,y,z)
,
\\
{\bf F}_{VSS}(x,y,z) &=& \frac{1}{x} \Bigl [-\lambda(x,y,z) {\bf I}(x, y, z) + (y-z)^2 {\bf I}(0, y, z)
+ x {\bf A}(y) {\bf A}(z) 
\nonumber \\ && 
+ \bigl \{2 x - y + z \bigr \} {\bf A}(x) {\bf A}(y)
+ \bigl \{2 x + y - z \bigr \} {\bf A}(x) {\bf A}(z)
\Bigr ]
,
\label{eq:fVSSxyzDRED}
\\
{\bf F}_{VSS}(0,x,y) &=& (d-1) \Bigl [(x + y)  {\bf I}(0, x, y) + {\bf A}(x) {\bf A}(y) \Bigr ] 
,
\\
{\bf F}_{VVS}(x,y,z) &=& \frac{1}{4 x y} \Bigl [ \bigl \{-12 x y -\lambda(x,y,z)  \bigr \} {\bf I}(x,y,z)
-z^2 {\bf I}(0,0,z) 
\nonumber \\ && 
+ (x-z)^2 {\bf I}(0,x,z) + (y-z)^2 {\bf I}(0,y,z) + (z - x - y) {\bf A}(x) {\bf A}(y) 
\nonumber \\ && 
+ y {\bf A}(x) {\bf A}(z) + x {\bf A}(y) {\bf A}(z) \Bigr ]
,
\\
{\bf F}_{VVS}(0,x,y) &=& \frac{1}{4x}\Bigl \{ \bigl [(d-13) x + (d-1) y \bigr ] {\bf I}(0,x,y) 
- (d-1) y {\bf I}(0,0,y)
\nonumber \\ && 
+ (d-1)  {\bf A}(x) {\bf A}(y) \Bigr \}
,
\\
{\bf F}_{VVS}(0,0,x) &=& \frac{1}{4} \left ( -d^2 + 5 d - 16 \right ) {\bf I}(0,0,x)
,
\label{eq:fVVS00xDRED}
\\
{\bf F}_{\rm gauge} (x,y,z) &=&  \frac{1}{4xyz} \Bigl [
\lambda(x,y,z) \bigl \{ -12 (x y + x z + y z) - \lambda(x,y,z)  \bigr \}  {\bf I}(x,y,z) 
\nonumber \\ && 
+ (x-y)^2 \bigl \{x^2 + y^2 + 10 x y \bigr \} {\bf I}(0,x,y) + z^2 (2 x y - z^2) {\bf I}(0,0,z)
\nonumber \\ && 
+ (x-z)^2 \bigl \{x^2 + z^2 + 10 x z \bigr \} {\bf I}(0,x,z) + y^2 (2 x z - y^2) {\bf I}(0,0,y)
\nonumber \\ && 
+ (y-z)^2 \bigl \{y^2 + z^2 + 10 y z \bigr \} {\bf I}(0,y,z) + x^2 (2 y z - x^2) {\bf I}(0,0,x)
\nonumber \\ && 
+ z \bigl \{ z^2 + 9 x (z-x) + 9 y (z - y) + 14 x y  \bigr \} {\bf A}(x) {\bf A}(y)
\nonumber \\ && 
+ y \bigl \{ y^2 + 9 x (y-x) + 9 z (y - z) + 14 x z \bigr \} {\bf A}(x) {\bf A}(z)
\nonumber \\ && 
+ x \bigl \{ x^2 + 9 y (x-y) + 9 z (x - z) + 14 y z \bigr \} {\bf A}(y) {\bf A}(z)
\Bigr ]
,
\\
{\bf F}_{\rm gauge} (0,x,y) &=&  
\frac{1}{4xy} \Bigl [
(x+y) \bigl \{ (d - 11)(x - y)^2 + 12 (d-1) x y  \bigr \} {\bf I}(0,x,y) 
\nonumber \\ && 
+ x^2 \bigl \{(11 - d) x + 2 y\bigr \}  {\bf I}(0,0,x) 
+ y^2 \bigl \{(11 - d) y + 2 x \bigr \}  {\bf I}(0,0,y)
\nonumber \\ && 
+ \bigl \{ (d-11) (x-y)^2 + 4 (3d-7) x y \bigr \} {\bf A}(x) {\bf A}(y) 
\Bigr ]
,
\\
{\bf F}_{\rm gauge} (0,0,x) &=& \frac{1}{4} (-32 + 25 d - d^2) x\hspace{1pt} {\bf I}(0,0,x)
.
\label{eq:Fgauge00x}
\eeq
For convenience, the dimensional reduction functions appearing in eqs.~(\ref{eq:FFFVxyz})-(\ref{eq:Fgauge00x}) are also given in the ancillary file {\tt bare2loop.txt} distributed with this paper. Note that these dimensional reduction results are valid only in models with softly broken supersymmetry. This is because in non-supersymmetric  models in which one might choose to use dimensional reduction as the regulator, the couplings of the $d$-dimensional vectors are not equal to those of the $(4-d)$-dimensional epsilon scalars \cite{Jack:1993ws,Jack:1994bn}.
 
%%%%%%%%%%%%%%%%%%%%%%%%%%%%%%%%%%%%%%%%%%%%%%%%%%%%%%%%%%%%%%%%%%%%%%%%%%%%%%%%%
\section{Rules for treating Goldstone bosons\label{sec:Goldstonerules}}
\setcounter{equation}{0}
\setcounter{figure}{0}
\setcounter{table}{0} 
\setcounter{footnote}{1}

The formulas in the preceding section for ${\bf f}_{SSS}$, ${\bf f}_{SS}$, ${\bf f}_{FFS}$, ${\bf f}_{\overline F  \overline F S}$,
${\bf f}_{VSS}$, and ${\bf f}_{VVS}$ (or ${\bf F}_{VSS}$ and ${\bf F}_{VVS}$ in the case of dimensional reduction) are valid for scalars that are not Goldstone bosons. Now consider what happens when we follow the prescription of eq.~(\ref{eq:expandGBprop}), in which perturbation theory is organized by treating the Goldstone-boson bare squared masses $G_B$  as interactions. The 
coefficients multiplying the master integral functions ${\bf A}$ and ${\bf I}$ are always polynomials
in the scalar masses, including $G_B$. Therefore, at two-loop order the impact of treating the Goldstone propagators as in eq.~(\ref{eq:expandGBprop}) can be summarized in terms of replacement rules for ${\bf A}$ and ${\bf I}$ 
when some of the squared-mass arguments are Goldstones.

For example, if at any loop order in the usual perturbation theory one encounters ${\bf A}(G_B)$, eq.~(\ref{eq:expandGBprop}) dictates that one should make the replacement
\beq
{\bf A}(G_B) &\rightarrow& \int_p \frac{1}{p^2} - G_B \int_p \frac{1}{(p^2)^2} + G_B^2 \int_p \frac{1}{(p^2)^3}
- G_B^3 \int_p \frac{1}{(p^2)^4} + \ldots,
\label{eq:AGBantiresummation}
\eeq
which vanishes trivially because each of the scale-free integrations vanishes identically in dimensional regularization. In particular, this means that with the prescription given here, the Goldstone boson one-loop diagrams make no contribution to the effective potential at all. This is consistent with the general idea that Goldstone-boson squared masses should be thought of as loop-suppressed, and with the conclusion arrived at within the resummation framework, where the one-loop Goldstone diagrams are resummed together with higher-order corrections. However, within the anti-resummation framework they are simply absent from the start.

More generally, when all of the propagators are massless particles or Goldstones, one has the simple replacement rules
\beq
{\bf A}(G_B) &\rightarrow& 0,
\label{eq:AGx}
\\
{\bf I}(0,0,G_B) &\rightarrow& 0,
\\
{\bf I}(0,G_{1B},G_{2B}) &\rightarrow& 0, 
\\
{\bf I}(G_{1B},G_{2B},G_{3B}) &\rightarrow& 0,
\eeq
where in the last two rules I have accounted for the possibility (not realized in the Standard Model) that there
may be Goldstone bosons with distinct squared masses.
It is also straightforward to apply eq.~(\ref{eq:expandGBprop}) to find the remaining 
replacement rules necessary for the bare two-loop effective potential:
\beq
{\bf I}(G_B,x,y) &\rightarrow&
{\bf I}(0,x,y) 
+ G_B \left[ 
\frac{2-d}{(x-y)^2} {\bf A}(x) {\bf A}(y) + \frac{(3-d)(x+y)}{(x-y)^2} {\bf I}(0,x,y) \right ] 
\nonumber \\ && 
%\!\!\!\!\!\!\!\!
+ 
G_B^2 \Bigl [\frac{(5-d)(2-d) (x+y)}{2(x-y)^4} {\bf A}(x) {\bf A}(y) 
\nonumber \\ && + \frac{(3-d)\left [(4-d) (x+y)^2 + 4 x y \right ]}{2 (x-y)^4} {\bf I}(0,x,y) \Bigr] 
%\nonumber \\ && 
+ \cdots
,
\label{eq:IGxy}
\\
{\bf I}(G_B,x,x) &\rightarrow&
\frac{(2-d)\left [{\bf A}(x) \right ]^2}{2 (d-3) x} 
\left [1 + G_B \frac{(3-d)}{2(5-d) x} + G_B^2 \frac{(4-d)(3-d)}{4 (7-d)(5-d) x^2}
+ \cdots
\right ] 
,
\phantom{xx}
\label{eq:IGxx}
\\
{\bf I}(0,G_B,x) &\rightarrow&
{\bf I}(0,0,x) \left [ 1 + G_B \left (\frac{3-d}{x} \right ) + G_B^2 \frac{(4-d)(3-d)}{2 x^2} + \cdots \right ]
,
\label{eq:I0Gx}
\\
{\bf I}(G_{1B},G_{2B},x) &\rightarrow&
{\bf I}(0,0,x) \biggr [ 1 + (G_{1B} + G_{2B}) \left (\frac{3-d}{x} \right ) 
+ (G_{1B}^2 + G_{2B}^2) \frac{(4-d)(3-d)}{2 x^2}
\nonumber \\ && 
+ G_{1B} G_{2B} \frac{(6-d)(3-d)}{x^2} + \cdots
\biggr ]
.
\label{eq:IG1G2x}
\eeq
Equations (\ref{eq:AGx})-(\ref{eq:IG1G2x}) can be substituted as needed into eqs.~(\ref{eq:fSSxy})-(\ref{eq:fVVSxyz}) and (\ref{eq:fVSS0xy})-(\ref{eq:fVVS00x}) and (\ref{eq:fVSSxyzDRED})-(\ref{eq:fVVS00xDRED}) in order to find the anti-resummation prescription
for Goldstone boson contributions to the bare two-loop effective potential. Note that I have included terms up
to second order in the bare Goldstone boson masses in eqs.~(\ref{eq:IGxy})-(\ref{eq:IG1G2x}). An examination of eqs.~(\ref{eq:deltaBnj}) and (\ref{eq:DeltaBonedeltaB})-(\ref{eq:DeltaBthreedeltaB}) reveals that this is the level necessary (along with three-loop master integral replacement rules up to linear order in $G$, to be given below) to have a complete three-loop result for the minimization of the effective potential.

It again bears emphasis that the arrows in eqs.~(\ref{eq:AGx})-(\ref{eq:IG1G2x}) certainly are {\em not} equalities. Instead, they denote the replacements that should be made in the usual perturbation theory to arrive at the corresponding formulas in the Goldstone boson anti-resummation version of bare perturbation theory promoted in the present paper. This should be obvious, since it is definitely not true that ${\bf A}(G_B) = 0$ or ${\bf I}(0,0,G_B)=0$ as functional identities. 

Related to this point, it is interesting to compare with the strategy of ref.~\cite{Espinosa:2017aew}, in which an approach to Goldstone boson resummation is presented that contains some similar formulas, but with a quite different interpretation and application. In that reference, motivated by the formalism of effective field theory, each Goldstone propagator is separated into hard-Goldstone ($G_h$) and soft-Goldstone ($G_s$) contributions. For example, one writes\footnote{This follows the notation of ref.~\cite{Espinosa:2017aew} although perhaps ${\bf I}(G, x, y) = {\bf I}_h(G, x, y) + {\bf I}_s(G, x, y)$ would be more literal.} ${\bf I}(G, x, y) = {\bf I}(G_h, x, y) + {\bf I}(G_s, x, y)$ as an actual functional equality, which is proved there. Now, the small-$G_h$ expansion of the hard Goldstone contribution ${\bf I}(G_h, x, y)$ in eq.~(108) of ref.~\cite{Espinosa:2017aew} has the identical form to the right side of my eq.~(\ref{eq:IGxy}) above. However, ${\bf I}(G_s, x, y)$ in eq.~(109) of ref.~\cite{Espinosa:2017aew} has no counterpart in the approach of the present paper. Similarly, eqs.~(61), (82), and (94) of ref.~\cite{Espinosa:2017aew} for hard Goldstone parts of integrals are identical in form to the right-hand sides of my eqs.~(\ref{eq:IG1G2x}), (\ref{eq:I0Gx}), and (\ref{eq:IGxx}), while the soft and mixed hard/soft Goldstone contributions found in ref.~\cite{Espinosa:2017aew} play no role and do not appear in the approach of the present paper. Nevertheless, after renormalization the end results are the same for the minimization of the renormalized effective potential, and for calculations of physical observables at a fixed loop order. The two approaches correspond to different organizations of perturbation theory. 

If the comparison between the two approaches seems perplexing, one may note that since logarithms of Goldstone-boson squared masses (which in ref.~\cite{Espinosa:2017aew} appear in the individual soft $G_s$ contributions after expansion in $\epsilon$ and derive from factors of ${\bf A}(G)$) must be absent in the end, it should not be too surprising that there exists an organization of perturbation theory in which they do not appear at all. Indeed, factors of ${\bf A}(G)$ clearly can never appear in the approach of the present paper. More concretely (as emphasized to me by an anonymous referee), 
the soft parts were shown in ref.~\cite{Espinosa:2017aew} to resum to terms proportional to $\overline G = G + \kappa \Delta^{(1)} + \kappa^2 \Delta^{(2)} + \kappa^3 \Delta^{(3)} + \ldots$, which vanishes at the minimum of the effective potential. This explains why the soft parts can be discarded in the minimization condition. Similar observations were made in ref.~\cite{Ekstedt:2018ftj}.  

For the master integrals for three-loop vacuum diagrams contributing to the effective potential, I follow the notations and conventions of ref.~\cite{Martin:2016bgz}. In particular, define
\beq
{\bf F}(w,x,y,z) &=&
\int_p \int_q \int_k \frac{1}{(p^2 + w)^2 (k^2 + x)(q^2 + y)[(p+k-q)^2 + z]}
,
\label{eq:defFF}
\\
{\bf G}(v,w,x,y,z) &=&
\int_p \int_q \int_k
\frac{1}{(p^2 + v) (k^2 + w) [(p-k)^2 + x](q^2 + y)[(q-p)^2 + z]}
,
\label{eq:defGG}
\\
{\bf H}(u,v,w,x,y,z) &=&
\int_p \int_q \int_k
\frac{1}{(p^2 + u)(k^2 + v)(q^2 + w) [(p-k)^2 + x][(k-q)^2 + y][(q-p)^2 + z]}
.
\nonumber \\ &&
\label{eq:defHH}
\eeq
When one or more of the propagators are Goldstones, one applies eq.~(\ref{eq:expandGBprop}), expanding to linear
order in the Goldstone-boson squared masses. Some of the results are quite lengthy, and there are a variety of special cases in which denominators in the coefficients multiplying the master integrals threaten to vanish for special configurations, which then had to be handled separately. Therefore, the resulting formulas will not be given explicitly in print here. Instead, the ancillary file {\tt GoldstoneRules.txt} gives the results for ${\bf I}(x,y,z)$, ${\bf F}(w,x,y,z)$, ${\bf G}(v,w,x,y,z)$, and ${\bf H}(u,v,w,x,y,z)$, for each possible combination of squared masses $u,v,w,x,y,z$ that are either 0, or Goldstones, or non-zero and not Goldstones. (The vanishing squared masses could belong e.g. to gauge bosons with unbroken symmetries, or to fermions protected by unbroken gauge or chiral symmetries.) It is important to note that one only needs the Goldstone rules for this finite number of master integrals. 

To illustrate, some relatively simple examples that involve only Goldstones and one other non-zero squared mass $x$ are
\beq
%{\bf F}(x, G_{1B}, G_{2B}, x) &=& {\bf F}(x,0,0,x) 
%+ 
%(G_{1B} + G_{2B}) \biggl [ 
%\frac{(d-2)^2}{24 x^3} {\bf A}(x)^3 
%\nonumber \\ &&
%+  \frac{1}{3} (d-4)(d-3) {\bf H}(0,0,x,0,x,x)
%\biggr ]
%,
%\\
{\bf F}(G_{1B}, G_{2B}, x, x) &\rightarrow& (3-d) {\bf G}(0,0,0,x,x) 
+
G_{1B}  \biggl [
\frac{(4-d) (d-2)^2}{48 (2d-9) x^3} {\bf A}(x)^3 
\nonumber \\ &&
+ \frac{(d-4)^2 (3-d)}{6 (2 d - 9)} {\bf H}(0,0,x,0,x,x)
\biggr ]
+
G_{2B}  \biggl [
\frac{(6-d) (d-2)^2}{24 (2d-9) x^3} {\bf A}(x)^3 
\nonumber \\ &&
+ \frac{(d-4) (d-3) (6-d)}{3 (2 d - 9)} {\bf H}(0,0,x,0,x,x)
\biggr ]
,
\\
{\bf G}(G_{1B}, G_{2B}, x, G_{3B}, x) &\rightarrow& {\bf G}(0,0,x,0,x) 
+ G_{1B} \biggl [ \frac{(3d-10)}{d x} {\bf G}(0,0,x,0,x)  
\nonumber \\ &&
+ \frac{2 (2-d)(d-3)}{d x^2} {\bf A}(x) {\bf I}(0,0,x) \biggr ]
\nonumber \\ &&
+ (G_{2B} + G_{3B}) \biggl [
\frac{4(d-4)(5-d)}{3(d-2)} {\bf H}(0,0,x,0,x,x)
\nonumber \\ &&
+ \frac{(5-d)(d-2)}{6 (d-3) x^3} {\bf A}(x)^3
+ \frac{(4-d)}{x^2} {\bf A}(x) {\bf I}(0,0,x) 
\biggr ]
,
\nonumber \\ &&
\\
{\bf H}(G_{1B}, G_{2B}, G_{3B}, x, x, x) &\rightarrow& {\bf H}(0,0,0,x,x,x) 
+ (G_{1B} + G_{2B} + G_{3B}) \biggl [
\frac{(d-3)}{x^2} {\bf G}(x,0,0,x,x)
\nonumber \\ &&
+ \frac{2(3-d)}{x^2} {\bf G}(0,0,0,x,x)
+ \frac{(4-d)}{2 x^2} {\bf G}(0,0,x,0,x) 
\nonumber \\ &&
+ \frac{(4-d)}{2 x} {\bf H}(0,0,0,x,x,x) 
\biggr ]
.
\eeq

The application of the prescription in eq.~(\ref{eq:expandGBprop}) to other calculations (beyond the effective potential) is straightforward.
For example, one-loop self-energy diagrams can always be reduced to the Passarino-Veltman integrals \cite{Passarino:1978jh} denoted here by
${\bf A}(x)$ and 
\beq
{\bf B}(x,y) &=& \int_k \frac{1}{(k^2 +x)[(k-p)^2 + y]}.
\eeq 
Then one-loop order self-energy functions
containing Goldstone bosons will make use of only the simple rules ${\bf A}(G_B) \rightarrow 0$ and
\beq
{\bf B}(G_B, x) &\rightarrow & {\bf B}(0,x) + G_B \left [ \frac{2-d}{(s-x)^2} {\bf A}(x) +
\frac{(3-d)(s+x)}{(s-x)^2} {\bf B}(0,x) \right ] 
\nonumber \\ && 
+ G_B^2 \left[ \frac{(5-d)(2-d)(s+x)}{2(s-x)^4} {\bf A}(x) +
\frac{(3-d)\left [(4-d)(s+x)^2 + 4 s x \right ]}{2(s-x)^4} {\bf B}(0,x) \right ]
\nonumber \\ && 
+ \ldots 
,
\\
{\bf B}(0,G_B) &\rightarrow & {\bf B}(0,0) \left [1 + G_B \left (\frac{3-d}{s}\right ) + G_B^2 \frac{(4-d)(3-d)}{2 s^2} + \cdots \right ]
,
\\
{\bf B}(G_{1B},G_{2B}) &\rightarrow & {\bf B}(0,0) \biggl [1 + (G_{1B} + G_{2B}) \left (\frac{3-d}{s}\right ) + (G_{1B}^2 + G_{2B}^2) 
\frac{(4-d)(3-d)}{2 s^2}
\nonumber \\ && 
+ G_{1B} G_{2B}  \frac{(6-d)(3-d)}{s^2}  + \cdots \biggr ]
,
\eeq
where $s = -p^2$. Going beyond one loop, it is clear that only non-negative integer powers of Goldstone-boson squared masses can occur. However, the present paper is not about self-energy functions, so details about the multi-loop self-energy case will be reserved to future work.

\section{Special three-loop master integral identities\label{sec:special}}
\setcounter{equation}{0}
\setcounter{figure}{0}
\setcounter{table}{0} 
\setcounter{footnote}{1}

The rules described in the previous section and given in {\tt GoldstoneRules.txt} necessarily involve non-generic configurations,
in which one or more of the squared masses vanish, and others may be equal to each other. In such cases, there are non-trivial identities between different three-loop candidate master integrals, which can be discovered using integration by 
parts.\footnote{The computer program {\tt KIRA} \cite{Maierhofer:2017gsa,Maierhofer:2018gpa,Klappert:2020nbg,Lange:2025fba} is one of several public codes that implement the integration-by-parts identities in an efficient way. It
was used to derive or check many of the results found in this paper.} For one example (of many), 
\beq
0 & = & 2 (x+y) {\bf F}(x,0,0,y) + (4-d) (x-y) {\bf G}(0,0,0,x,y) + (d-2) y {\bf G}(0,0,x,0,y) 
\nonumber \\
&& + (d-2) {\bf A}(y) {\bf I}(0,0,x) + (d-2) y {\bf A}(x) {\bf I}(0,0,y)/x. 
\label{eq:FFx00yid}
\eeq
A complete list of the useful identities of this type that I know of for non-generic squared-mass arguments in three-loop vacuum master integrals is given in the ancillary file {\tt identities.txt} in a form suitable for use in computer programs. It extends a similar ancillary file with the same name that was provided with ref.~\cite{Martin:2025cas}.

Now, when using such identities to eliminate integrals from the list of masters, before or after applying the rules of the ancillary file {\tt GoldstoneRules.txt}, it is important to avoid explicit factors of $(4-d)$ in denominators, so as to have an $\epsilon$-finite basis in the sense of ref.~\cite{Chetyrkin:2006dh,Martin:2021pnd}. For example, when using eq.~(\ref{eq:FFx00yid}), one must avoid solving for and eliminating ${\bf G}(0,0,0,x,y)$, and instead choose to eliminate either ${\bf F}(x,0,0,y)$ or
${\bf G}(0,0,x,0,y)$.
In almost all special cases of squared mass arguments, a subset of the master integrals ${\bf A}$, ${\bf I}$, ${\bf F}$, ${\bf G}$ and ${\bf H}$ can be chosen to form an $\epsilon$-finite basis. For three-loop vacuum diagrams, there are only two special cases that (as far as I know) require special treatment by enlarging the set of candidate master integrals to include special cases of 
\beq
{\bf H}'(u,v,w,x,y,z) &\equiv& \frac{\partial }{\partial u} {\bf H}(u,v,w,x,y,z) .
\eeq
(The prime denotes the partial derivative with respect to the first squared mass argument.) Note that as $\epsilon \rightarrow 0$, the ${\bf H}'(u,v,w,x,y,z)$ integral is always free of ultraviolet (UV) poles (by power counting), and is free of IR poles provided that $u \not= 0$ (due to having no doubled massless propagators).

The first special case is that to avoid explicit factors of $1/\epsilon$ in the bare effective potential formulas it is sometimes necessary to eliminate ${\bf G} (0, 0, x, 0, y)$ using the following identity:
\beq
{\bf G} (0, 0, x, 0, y) &=& 
\frac{(2-d)(3d-10)(x+y)}{4 (d-3)^2 x y (x-y)} {\bf A}(x)^2 {\bf A}(y)
+ \frac{(3d-10)(x+y)}{(d-3)(d-2)(x-y)} {\bf F}(y,x,x,y)
\nonumber \\ &&
+ \frac{(4-d)(5d-16)}{(d-3)(d-2)} {\bf G}(0,0,0,x,y)
+ \frac{(4-d)(3d-10)x}{(d-3)(d-2)(x-y)} {\bf G}(0,x,x,y,y)
\nonumber \\ &&
+ \frac{(4-d)[(d-5)x + (7 - 2 d) y]}{(d-3)(d-2)} {\bf H}(0,x,x,y,0,y)
- {\bf A}(x) {\bf I}(0,0,y)/x 
\nonumber \\ &&
- {\bf A}(y) {\bf I}(0,0,x)/y 
+ \frac{(2d-7) x + (d-3) y}{(3-d) x (x - y)} {\bf A}(x) {\bf I}(0,x,y)
\nonumber \\ &&
- \frac{(d-3) x + (2d-7) y}{(3-d) y (x - y)} {\bf A}(y) {\bf I}(0,x,y)
\nonumber \\ &&
+ \frac{4 (d-4) x (x-y)}{(d-3)(d-2)} {\bf H}'(x,0,x,y,y,0) .
\label{eq:GG00x0y}
\eeq
One can see from this identity that if one instead attempted to eliminate ${\bf H}'(x,0,x,y,y,0)$ as usual, there would result explicit poles $1/(d-4)$.
It is not always necessary to apply this identity, as some expressions involving ${\bf G} (0, 0, x, 0, y)$ are already free of explicit poles in $d-4$ without its use. Furthermore, it is never necessary to apply it if $x=y$ or $x=0$ or $y=0$, because in those cases the simpler identities 
\beq
{\bf G} (0, 0, x, 0, x) &=& 
\frac{16 (4-d)(2d-7)x}{3(d-2)(3d-10)} {\bf H}(0,0,x,0,x,x) 
- 2 {\bf A}(x) {\bf I}(0,0,x)/x
\nonumber \\ &&
+ \frac{2(2-d)(2d-7)}{3(d-3)(3d-10) x^2} {\bf A}(x)^3
,
\\
{\bf G} (0, 0, 0, 0, x) &=& 
\frac{(d-4) x}{3d-10} {\bf H}(0,0,0,0,x,x)
+ \frac{4 (4-d) x}{3(3d-10)} {\bf H}(0,0,x,0,x,x)
\nonumber \\ &&
+ \frac{(2-d)}{(3d-10)x} {\bf A}(x) {\bf I}(0,0,x)
+ \frac{(d-2)^2}{6(3-d)(3d-10) x^2} {\bf A}(x)^3
\eeq
are sufficient.

Another important requirement is that one should not tolerate ${\bf F}$ as one of the master integrals when its first argument vanishes exactly.\footnote{An alternative approach replaces a vanishing first argument of ${\bf F}$ by an IR-regulator squared mass $w$, and then takes the limit $w \rightarrow 0$ only after expanding in $\epsilon$
and combining contributions from all diagrams. This was the method used in ref.~\cite{Martin:2017lqn}, for example. However, this procedural option cannot be used with the anti-resummation treatment of the present paper if the first argument came from a Goldstone-boson squared mass, because eq.~(\ref{eq:expandGBprop}), and the rules given in section \ref{sec:Goldstonerules} that follow from it, require that the Goldstone boson propagators are exactly massless.} This is because ${\bf F}(0,x,y,z)$ has an IR divergence due to the doubled massless propagator, resulting in a problematic expansion in $\epsilon$. Fortunately, it is always possible to eliminate such integrals in favor
of master integrals that are not IR divergent. This is very easy when one of the other arguments also vanishes, due to the simple identity
\beq
{\bf F}(0,0,x,y) &=& (3-d) {\bf G}(0,0,0,x,y) .
\eeq
When only the first argument vanishes, one way to proceed is to make use of the identity  
\beq
{\bf F}(0,x,y,z) &=& \frac{1}{2z} \biggl [
(x-y) {\bf F}(y,x,x,y) + \frac{1}{2} \left [(4-d)(x-y) + (3d-10) z \right ] {\bf G}(0,x,x,y,y)
\nonumber \\ &&
+  (3-d)(y-x+z) {\bf G}(x,0,x,y,z)
+  (3-d)(x-y+z) {\bf G}(y,0,y,x,z)
\nonumber \\ &&
+ \frac{1}{2} (d-4) \left [(x-y)^2 + 2 (x+y) z - 3 z^2 \right ] {\bf H}(0,x,y,x,z,y)
\nonumber \\ &&
+ \lambda(x,y,z) z {\bf H}'(z,x,x,y,0,y)
+ \frac{(d-2)^2 (x-y)}{4 (3-d) x y} {\bf A}(x)^2 {\bf A}(y)
\biggr ]
.
\label{eq:FF0xyzspecial}
\eeq
This generalizes eq.~(2.24) of ref.~\cite{Martin:2025cas}.

The expansions in small $\epsilon$ for ${\bf F}$, ${\bf G}$, and ${\bf H}$ involve renormalized master integrals $F(w,x,y,z)$, $\overline F(0,x,y,z)$, $G(v,w,x,y,z)$, $H(u,v,w,x,y,z)$ defined in ref.~\cite{Martin:2016bgz}, which also provides an algorithm and a public code for their numerical evaluation in general.
However, after using eqs.~(\ref{eq:GG00x0y}) and (\ref{eq:FF0xyzspecial}), one also needs the $\epsilon \rightarrow 0$ limit of the special case ${\bf H}'$ master integrals. These can be found by first computing the corresponding renormalized master integrals $H'(x,0,x,y,y,u)$ and $H'(z,x,x,y,u,y)$, using the results in the ancillary file {\tt derivatives.txt} distributed with ref.~\cite{Martin:2016bgz}, and then taking the limit $u \rightarrow 0$. Explicitly, the results needed for expanding eqs.~(\ref{eq:GG00x0y}) and (\ref{eq:FF0xyzspecial}) in $\epsilon$ are  
\beq
{\bf H}'(x,0,x,y,y,0) &=&
\frac{1}{x(x-y)} \Bigl [
\frac{1}{4} G(0,x,x,y,y)
+ \frac{1}{2} G(0,0,0,x,y) 
- \frac{1}{2} G(x,0,x,0,y) 
\nonumber \\ &&
+ \frac{1}{4} F(x,x,y,y) 
- \frac{1}{4 x y} A(x) A(y)^2
+ \frac{1}{2 x} A(x) A(y)
+ \frac{1}{4 y} A(y)^2
\nonumber \\ &&
- \frac{9x + 8 y}{16 x} A(x) 
- A(y) + x + \frac{7y}{6} - 2 x \zeta_3
\Bigr ] + {\cal O}(\epsilon)
,
\label{eq:Hpx0xyy0}
\\
{\bf H}'(z,x,x,y,0,y) &=& \frac{1}{\left [\lambda(x,y,z) \right ]^2}
\Bigl \{ 
2 x (x-y-z) G(x, 0, x, y, z) 
+ 2 y (y -x - z) G(y, 0, y, x, z)
\nonumber \\ &&
+ (z - x + y)(z + x - y) G(z,x,y,x,y)
- \lambda(x,y,z) G(0, x, x, y, y)
\nonumber \\ &&
+ 2 x (z - x + y) F(x, 0, y, z) 
+ 2 y (z + x - y) F(y, 0, x, z)
\nonumber \\ &&
+ 2 (x - y - z) (x - y + z) F(z, 0, x, y) 
+ 2 \lambda(x,y,z) \overline{F} (0, x, y, z)
\nonumber \\ &&
+ 2 x (z - x + y) F(x, x, y, y) 
+ 2 y (z + x - y) F(y, x, x, y)
+ 2 \bigl [3 (x + y) z 
\nonumber \\ && 
- (x - y)^2 - 2 z^2 
+ (z - x + y) A(x) 
+ (z + x - y) A(y) 
\bigr ] I(x,y,z)
\nonumber \\ && 
+ (x - y + z) \left [z - x - y + 2 A(y) + (x - y - z) A(z)/z  \right ] I(0,x,x)
\nonumber \\ && 
+ (y - x + z) \left [z - x - y + 2 A(x) + (y - x - z) A(z)/z  \right ] I(0,y,y)
\nonumber \\ && 
+ (x^2 - 3 x y + 2 y^2 - x z - 2 y z) A(x)
+ (y^2 - 3 x y + 2 x^2 - y z - 2 x z) A(y)
\nonumber \\ && 
+ \frac{1}{2} \left [4 (x+y) z - z^2 - 3 (x-y)^2 \right ] A(z)
+ 4 (x + y - z) \lambda(x,y,z) \zeta_3
\nonumber \\ && 
+ \frac{2}{3} (x+y-z) [z^2 + 5 (x+y) z - 6 (x-y)^2]
\Bigr \}
+ {\cal O}(\epsilon)
.
\label{eq:Hpzxxy0yexp}
\eeq
Equation (\ref{eq:Hpzxxy0yexp}) generalizes eq.~(2.27) of ref.~\cite{Martin:2025cas}. For convenience,
eqs.~(\ref{eq:Hpx0xyy0}) and (\ref{eq:Hpzxxy0yexp}) are implemented in the ancillary file {\tt HHpexp.txt} distributed with the present paper.

\section{Application to the Standard Model\label{sec:StandardModel}}
\setcounter{equation}{0}
\setcounter{figure}{0}
\setcounter{table}{0} 
\setcounter{footnote}{1}

Consider the Standard Model, with the bare tree-level potential as given in eq.~(\ref{eq:Vbare0SM}). In the following, I neglect all Yukawa couplings except that of the top quark. The bare field-dependent squared masses of the Goldstone bosons, the top quark, and the $W$ and $Z$ bosons are denoted by
\beq
G_B &=& \frac{1}{\phi_B} \frac{\partial V^{(0)}_B }{\partial \phi_B} \,=\, m_B^2 + \lambda_B \phi_B^2 ,
\\
t_B &=& \frac{1}{2} y_{tB}^2 \phi_B^2,
\\
W_B &=& \frac{1}{4} g_B^2 \phi_B^2, 
\\
Z_B &=& \frac{1}{4} (g_B^2 + g_B^{\prime 2}) \phi_B^2. 
\eeq
For the Higgs particle, there are two different choices for parameterizing the bare field-dependent squared mass,
which I call $H_B$ and $h_B$. They are defined as 
\beq
H_B &=& \frac{\partial^2 V^{(0)}_B }{\partial \phi_B^2} \,=\, m_B^2 + 3 \lambda_B \phi_B^2 
,
\\
h_B &=& H_B - G_B \,=\, 2 \lambda_B \phi_B^2 .
\eeq
Since $H_B$ depends on $m_B^2$, it is useful to eliminate it everywhere, and write all expressions in terms of $h_B$ instead. This means that the one-loop effective potential does have a dependence on $G_B$, due to the straightforward and non-singular power series expansion
\beq
{\bf f}(H_B) = {\bf f}(h_B) + G_B  {\bf f}'(h_B) + \frac{1}{2} G_B^2 {\bf f}''(h_B) + \frac{1}{6} G_B^3 {\bf f}'''(h_B) + \cdots .
\eeq
More generally, in all loop diagrams one applies the replacement rule
\beq
\frac{1}{p^2 + H_B} & \rightarrow & \frac{1}{p^2 + h_B} - \frac{G_B}{(p^2 + h_B)^2} + \frac{G_B^2}{(p^2 + h_B)^3} 
- \frac{G_B^3}{(p^2 + h_B)^4} + \cdots
.
\label{eq:HBhBexpansion}
\eeq 
Also, applying eq.~(\ref{eq:AGx}), we see that ${\bf f}(G_B)$ in the usual perturbative expansion is replaced by $0$ in the prescription of this paper. 
It follows that the one-loop contribution to the bare effective potential is
\beq
V^{(1)}_B &=& \frac{1}{d} \Bigl [(d-1) Z_B {\bf A}(Z_B) + 2 (d-1) W_B {\bf A}(W_B)  - 4 N_c t_B {\bf A}(t_B) + H_B {\bf A}(H_B)\Bigr ] ,
\label{eq:V1SMpreHexp}
\eeq
or, applying eq.~(\ref{eq:HBhBexpansion}),
\beq
V^{(1)}_B &=& \frac{1}{d} \Bigl [(d-1) Z_B {\bf A}(Z_B) + 2 (d-1) W_B {\bf A}(W_B)  - 4 N_c t_B {\bf A}(t_B) +   h_B {\bf A}(h_B)\Bigr ] 
\nonumber \\ && 
+ {\bf A}(h_B) \left [\frac{1}{2}  G_B + \Bigl (\frac{d-2}{8h_B}\Bigr ) G_B^2 + \frac{(d-4)(d-2)}{48 h_B^2} G_B^3 + \ldots \right ] .
\eeq

The two-loop bare contribution at second order in $G_B$, 
\beq
V^{(2)}_B &=& V^{(2,0)}_B + G_B V^{(2,1)}_B + G_B^2 V^{(2,2)}_B + \ldots
,
\eeq
is obtained by applying eqs.~(\ref{eq:AGx})-(\ref{eq:IG1G2x}) to all Goldstone boson propagators and eq.~(\ref{eq:HBhBexpansion}) to all Higgs boson propagators. The $G_B$-independent term is\footnote{For mere typographical convenience, I omit the $B$ subscript in the next six equations (\ref{eq:VB20SM})-(\ref{eq:DeltaB2SM}), with the understanding that all parameters $G$, $\phi$, $\lambda$, $g$, $g'$, $g_3$, $y_t$, $h$, $t$, $W$, and $Z$ appearing in them are actually the bare ones.}
\beq
V^{(2,0)}_B &=& 
\frac{3}{4} \lambda \hspace{1pt} {\bf f}_{SS}(h,h)
+ \frac{3}{2} \lambda h \bigl [ {\bf f}_{SSS}(h,h,h) + {\bf f}_{SSS}(0,0,h) \bigr ]
\nonumber \\ && 
+ \frac{1}{2} N_c y_t^2 \bigl [ {\bf f}_{FFS}(t,t,h) + t\hspace{1pt} {\bf f}_{\overline F \overline FS}(t,t,h) 
+ {\bf f}_{FFS}(t,t,0) - t\hspace{1pt} {\bf f}_{\overline F \overline FS}(t,t,0) 
+ 2\hspace{1pt} {\bf f}_{FFS}(0,t,0) \bigr ]
\nonumber \\ && -N_c \left ( g_3^2 C_F  + \frac{4}{9} \frac{g^2 g^{\prime 2}}{g^2 + g^{\prime 2}}\right ) 
t \hspace{1pt} {\bf f}_{\overline F \overline FV}(t,t,0) 
\nonumber \\ && 
+ \frac{1}{2} N_c \left [ (a_{u_L}^2 + a_{u_R}^2)\hspace{1pt} {\bf f}_{FFV}(t,t,Z) 
- 2 a_{u_L} a_{u_R} t \hspace{1pt} {\bf f}_{\overline F\overline FV}(t,t,Z) \right ]
\nonumber \\ && 
+ \frac{1}{2} \Bigl \{ N_c (n_G - 1) (a_{u_L}^2 + a_{u_R}^2) + N_c n_G (a_{d_L}^2 + a_{d_R}^2)
+ n_G (a_{e_L}^2 + a_{e_R}^2 + a_{\nu_L}^2) 
\Bigr  \}\hspace{1pt} {\bf f}_{FFV}(0,0,Z) 
\nonumber \\ && 
+ \frac{1}{2} g^2 N_c\hspace{1pt} {\bf f}_{FFV}(0,t,W) 
+ \frac{1}{2} g^2 \left \{N_c (n_G - 1) + n_G \right \}\hspace{1pt} {\bf f}_{FFV}(0,0,W)
\nonumber \\ && 
+ \frac{1}{8} (g^2 + g^{\prime 2})\hspace{1pt} {\bf f}_{VSS}(Z,0,h)
+ \frac{(g^2 - g^{\prime 2})^2}{8 (g^2 + g^{\prime 2})}\hspace{1pt} {\bf f}_{VSS}(Z,0,0)
+ \frac{1}{4} g^2 \hspace{1pt} {\bf f}_{VSS}(W,0,h) 
\nonumber \\ && 
+ \frac{1}{4} g^2 \hspace{1pt} {\bf f}_{VSS}(W,0,0)
+ \frac{1}{4} (g^2 + g^{\prime 2}) Z\hspace{1pt} {\bf f}_{VVS}(Z,Z,h)
+ \frac{1}{2} g^2 W\hspace{1pt} {\bf f}_{VVS}(W,W,h)
\nonumber \\ && 
+ \frac{g^{\prime 4}}{(g^2 + g^{\prime 2})} W\hspace{1pt} {\bf f}_{VVS}(W,Z,0)
+ \frac{g^2 g^{\prime 2}}{(g^2 + g^{\prime 2})} W\hspace{1pt} {\bf f}_{VVS}(0,W,0)
\nonumber \\ &&
+ \frac{g^2 g^{\prime 2}}{2(g^2 + g^{\prime 2})}\hspace{1pt} {\bf f}_{\rm gauge}(W,W,0)
+ \frac{g^4}{2(g^2 + g^{\prime 2})}\hspace{1pt} {\bf f}_{\rm gauge}(W,W,Z) ,
\label{eq:VB20SM}
\eeq
in which $N_c = 3$ is the number of colors, $C_F = (N_c^2 - 1)/2 N_c = 4/3$, and $n_G = 3$ is the number of fermion generations, and the couplings of the $Z$ boson to Standard Model fermions are denoted
\beq
a_f = \sqrt{g^2 + g^{\prime 2}} \left ( I_3^{f} - Q_f \frac{g^{\prime 2}}{g^2 + g^{\prime 2}} \right ),
\eeq
where $I_3^{u_L} = I_3^{\nu_L} = 1/2$, and $I_3^{d_L} = I_3^{e_L} = -1/2$, and 
$I_3^{u_R} = I_3^{d_R} = I_3^{e_R} = 0$, and $Q_{u_L} = Q_{u_R} = 2/3$, and $Q_{d_L} = Q_{d_R} = -1/3$
and $Q_{e_L} = Q_{e_R} = -1$, and $Q_{\nu_L} = 0$.

There are cancellations between different parts of eq.~(\ref{eq:VB20SM}), due to the fact that the would-be Goldstone bosons correspond to the longitudinal components of the vector fields $W$ and $Z$. For example,
the terms involving ${\bf f}_{VSS}(W,0,0)$, ${\bf f}_{VVS}(W,Z,0)$, ${\bf f}_{VVS}(0,W,0)$, ${\bf f}_{\rm gauge}(W,W,0)$, and ${\bf f}_{\rm gauge}(W,W,Z)$ each contain contributions proportional to ${\bf I}(0,0,W)$. However,
in their sum, the ${\bf I}(0,0,W)$ contributions completely cancel. This exact cancellation comes about only because the Goldstone-boson propagator squared mass is exactly 0 in the organization
of perturbation theory used here. [There remains a contribution proportional to $(N_c (n_G - 1) + n_G) {\bf I}(0,0,W)$ from the ${\bf f}_{FFV}(0,0,W)$ term involving light quark and lepton loops.] Similar cancellations occur for terms proportional to each of ${\bf I}(0,0,Z)$, and ${\bf I}(0,W,Z)$, and ${\bf I}(0,0,h)$, and ${\bf I}(0,h,W)$, and ${\bf I}(0,h,Z)$, and ${\bf I}(0,0,t)$. These cancellations are manifest in the expanded form of $V^{(2,0)}_B$, which is provided in an ancillary file {\tt V2SMbare.txt}.

The contributions linear and quadratic in $G_B$ are
\beq
V^{(2,1)}_B
&=& \frac{3 }{4h}(d-2) \lambda {\bf A}(h)^2 
- \frac{3}{2} (d-3) \lambda {\bf I}(h,h,h) -3 \lambda  {\bf I}(0,0,h) 
\nonumber \\ &&
+ \frac{N_c y_t^2}{2h} \bigl [ (2 d - 6) t + (2-d) h \bigr ] {\bf I}(h,t,t)
+ N_c y_t^2 \frac{(d-2)}{4 (d-3) h t} \bigl [h + (2d-6) t \bigr ] {\bf A}(t)^2
\nonumber \\ &&
+ N_c y_t^2 (2-d) \bigl [{\bf I}(0,0,t) + {\bf A}(t) {\bf A}(h)/h \bigr ]
\nonumber \\ &&
+ (g^2 + g^{\prime 2}) \frac{(d-1)(h - 2 Z) }{16 h Z (4 Z - h)} \bigl [h^2 - 4 h Z + 4 (d-3) Z^2 \bigr ]{\bf I}(h,Z,Z)
\nonumber \\ &&
+ g^2 \frac{(d-1)(h - 2 W) }{8 h W (4 W - h)} \bigl [h^2 - 4 h W + 4 (d-3) W^2 \bigr ] {\bf I}(h,W,W)
\nonumber \\ && 
+ (g^2 + g^{\prime 2}) \frac{(d-1)(h + Z)}{8 Z} 
{\bf I}(0,h,Z)
+ g^2 \frac{(d-1)(h + W)}{4 W} {\bf I}(0,h,W)
\nonumber \\ && 
+ (g^2 + g^{\prime 2}) \frac{(d-2) (d-1) (h - 8 Z)}{16 h (h - 4 Z)} {\bf A}(h) {\bf A}(Z)
+ g^2 \frac{(d-2) (d-1) (h - 8 W)}{8 h (h - 4 W)} {\bf A}(h) {\bf A}(W)
\nonumber \\ && 
+ \frac{g^{\prime 4}}{(g^2 + g^{\prime 2})} \frac{(d-1)(W+Z)}{4 Z (Z-W)^2} [W^2 + (4d - 14) W Z + Z^2]
{\bf I}(0,W,Z)
\nonumber \\ && 
+ g^2 \frac{(d-1)}{4 Z^2} [(3d-7) Z^2 + (10 - 3 d) W Z - W^2] {\bf I}(0,0,W)
+ g^2 \frac{(d-1)(3W - 2 Z)}{4 Z} {\bf I}(0,0,Z)
\nonumber \\ && 
+ (g^2 + g^{\prime 2}) \frac{(d-1)}{16 h Z (h - 4 Z)} [h^2 - 4 h Z + (4d-8) Z^2] {\bf A}(Z)^2
\nonumber \\ && 
+ g^2 \frac{(d-1)}{8 h W (h - 4 W)} [h^2 - 4 h W + (4d-8) W^2] {\bf A}(W)^2
\nonumber \\ && 
+ g^2 \frac{(d-1)}{4 W Z^2} [W^2 + (4 d- 10) W Z + Z^2] {\bf A}(W) {\bf A}(Z)
,
\eeq
and
\beq
V^{(2,2)}_B
&=& 
\frac{3 (d-3) (d-2)}{8h^2} \lambda {\bf A}(h)^2 
- \frac{3(d-4)(d-3)}{4h} \lambda {\bf I}(h,h,h) 
-\frac{3 (d^2 - 7 d + 13)}{2 h} \lambda  {\bf I}(0,0,h) 
\nonumber \\ &&
+ N_c y_t^2 \frac{(d-3)}{4h^2 (4t-h)} [ 4 (d-5) t^2 - 4 (d-3) h t + (d-2) h^2] {\bf I}(h,t,t)
\nonumber \\ &&
+ N_c y_t^2 \frac{(d-2)}{8 (d-5) h^2 t^2 (4t - h)} [4 (d-5)^2 t^3 - 2 (d-3)(d-5) h t^2 + 4 h^2 t - h^3]
{\bf A}(t)^2
\nonumber \\ &&
+ N_c y_t^2 \frac{(d-3)(d-2)}{2t} {\bf I}(0,0,t) 
+ N_c y_t^2 \frac{(d-2)}{2 h^2 (4t-h)} [(11 - 3d) t + h (d-3)] {\bf A}(t) {\bf A}(h) 
\nonumber \\ &&
+ (g^2 + g^{\prime 2}) \frac{(d-1)}{32 h^2 Z (h - 4 Z)^2} [(2-d) h^4 + 8 (d-2) h^3 Z - 4 (d-1)^2 h^2 Z^2 
\nonumber \\ && \qquad 
+ 16 (d-3)^2 h Z^3 - 16 (d-5)(d-3) Z^4]{\bf I}(h,Z,Z)
\nonumber \\ &&
+ g^2 \frac{(d-1)}{16 h^2 W (h - 4 W)^2} [(2-d) h^4 + 8 (d-2) h^3 W - 4 (d-1)^2 h^2 W^2 
\nonumber \\ && \qquad 
+ 16 (d-3)^2 h W^3 - 16 (d-5)(d-3) W^4]{\bf I}(h,W,W)
\nonumber \\ && 
+ (g^2 + g^{\prime 2}) \frac{(d-1)}{16 Z (h - Z)^2} [(d-2) h^2 + (4 - 2 d) h Z + (10 - 3 d) Z^2] {\bf I}(0,h,Z)
\nonumber \\ && 
+ g^2 \frac{(d-1)}{8 W (h - W)^2} [(d-2) h^2 + (4 - 2 d) h W + (10 - 3 d) W^2] {\bf I}(0,h,W)
\nonumber \\ && 
+ (g^2 + g^{\prime 2}) \frac{(d-2) (d-1)}{64 h^2 (h - Z)^2 (h - 4 Z)^2}  
\bigl [(d-12) h^4 + (140 - 22 d) h^3 Z 
\nonumber \\ && \qquad 
+ (89d - 408) h^2 Z^2 + (384 - 116 d) h Z^3 +  (48d - 176) Z^4 \bigr ]
{\bf A}(h) {\bf A}(Z)
\nonumber \\ && 
+ g^2 \frac{(d-2) (d-1)}{32 h^2 (h - 4 W)^2 (h - W)^2} \bigl [ (d-12) h^4 + (140 - 22 d) h^3 W 
\nonumber \\ && \qquad 
+ (89d - 408) h^2 W^2 + (384 - 116 d) h W^3 + (48 d - 176) W^4 
\bigr ]
{\bf A}(h) {\bf A}(W)
\nonumber \\ && 
+ g^2 \frac{(d-1)}{8 W Z^2 (Z-W)^2} \bigl [ 
(2 - d) (W^4 + Z^4) + (-4 d^2 + 28 d - 44) (W^3 Z + W Z^3) 
\nonumber \\ && \qquad 
+ (d-6)(26 - 8 d) W^2 Z^2 
\bigr ]
{\bf I}(0,W,Z)
\nonumber \\ && 
+ g^2 \frac{(d-1)}{8 W Z^2} [(d-2) W^2 + (3 d^2 - 21 d + 34) W Z + (3d - 10)(2 - d) Z^2] {\bf I}(0,0,W)
\nonumber \\ && 
+ g^2 \frac{(d-1)}{8 W Z^2} [(22 - 7 d) W^2 + (6d - 20) W Z + (4-d) Z^2 ] {\bf I}(0,0,Z)
\nonumber \\ && 
+ (g^2 + g^{\prime 2}) \frac{(d-2)(d-1)(h - 2 Z)}{32 h^2 Z (h - 4 Z)^2} [h^2 - 4 h Z + (4d-20) Z^2] {\bf A}(Z)^2
\nonumber \\ && 
+ g^2 \frac{(d-2)(d-1)(h - 2 W)}{16 h^2 W (h - 4 W)^2} [h^2 - 4 h W + (4d-20) W^2] {\bf A}(W)^2
\nonumber \\ && 
- g^2 \frac{(d-2)(d-1)(W+Z)}{8 W Z^2 (Z - W)^2} [W^2 + (4 d- 22) W Z + Z^2] {\bf A}(W) {\bf A}(Z)
.
\eeq
For convenience, these expressions for $V^{(2,1)}_B$ and $V^{(2,2)}_B$ are given along with $V^{(2,0)}_B$ in
the ancillary file {\tt V2SMbare.txt}.

Now, applying eqs.~(\ref{eq:DeltaBonedeltaB}) and (\ref{eq:DeltaBtwodeltaB}), with eq.~(\ref{eq:deltaBnj}), one obtains the results for the minimization of the two-loop bare effective potential,
\beq
\Delta_B^{(1)} &=&
3 \lambda {\bf A}(h) - 2 N_c y_t^2 {\bf A}(t) + \frac{1}{2} g^2 (d-1) {\bf A}(W) + \frac{1}{4} (g^2 + g^{\prime 2}) (d-1) {\bf A}(Z),
\\
\Delta_B^{(2)} &=& \frac{1}{\phi^2}\biggl [
N_c \left ( g_3^2 C_F  + \frac{4}{9} \frac{g^2 g^{\prime 2}}{g^2 + g^{\prime 2}}\right )
\frac{2 (d-1)(d-2)^2}{d-3} {\bf A}(t)^2 
\nonumber \\ &&
+ N_c (a_{u_L}^2 + a_{u_R}^2) (d-2) \Bigl \{ \bigl [ (2d-6) t + (2-d) Z \bigr ] {\bf I}(t,t,Z) 
+ (4 - 2 d) {\bf A}(t) {\bf A}(Z)
\nonumber \\ &&   + (d-2) {\bf A}(t)^2 
\Bigr \}
- 4 N_c a_{u_L} a_{u_R} (d-2)(d-1) t \hspace{1pt} {\bf I}(t,t,Z)
- \Bigl \{ N_c (n_G - 1) (a_{u_L}^2 + a_{u_R}^2) \nonumber \\ &&
+ N_c n_G (a_{d_L}^2 + a_{d_R}^2)
+ n_G (a_{e_L}^2 + a_{e_R}^2 + a_{\nu_L}^2) 
\Bigr  \} (d-2)^2 Z \hspace{1pt} {\bf I}(0,0,Z)
\nonumber \\ &&
+ N_c g^2 \frac{(d-2)}{W}  \Bigl \{ (t-W) \bigl [t + (d-2) W \bigr ] {\bf I}(0,t,W) 
+ \bigl [t + (2-d) W \bigr ] {\bf A}(W) {\bf A}(t)
\nonumber \\ &&
- t^2 {\bf I}(0,0,t)  \Bigr \}
- \bigl [N_c (n_G - 1) + n_G \bigr ] g^2 (d-2)^2 W {\bf I}(0,0,W)
+ \frac{1}{2} \lambda (21 - 9 d) h \hspace{1pt} {\bf I}(h,h,h) 
\nonumber \\ &&
- 3 \lambda h {\bf I}(0,0,h)
+ 2 (d-1) \lambda (h + W) {\bf I}(0,h,W)
+ (d-1) \lambda (h+Z) {\bf I}(0,h,Z)
\nonumber \\ && 
+ \frac{1}{2} N_c y_t^2 \bigl [ (6 - 3 d) h + (10 d - 22) t \bigr ] {\bf I}(h,t,t)
+ N_c y_t^2 (2 - d) (h - 2 t) {\bf I}(0,0,t)
\nonumber \\ &&
+ \frac{1}{2} N_c y_t^2 (6 - 3d) {\bf A}(h) {\bf A}(t) 
+ N_c y_t^2 \frac{(d-2)}{4 (d-3) t} \bigl [h + 10 (d-3) t \bigr ] {\bf A}(t)^2
\nonumber \\ &&
+ (3 g^2 - g^{\prime 2}) (d-2) \bigl [(d-1) W^2/Z + (2d - 3) W  + Z/4 \bigr ] {\bf I}(W,W,Z)
\nonumber \\ && 
+ \frac{g^2}{8 W (h - 4 W)} \Bigl \{ (5 - 3 d) h^3 + (22d - 38) h^2 W + (44 - 12 d^2) h W^2 
\nonumber \\ &&
+ 8 (d-1)(5d-11) W^3 \Bigr \} {\bf I}(h,W,W)
+ \frac{g^2 + g^{\prime 2}}{16 Z (h - 4 Z)} \Bigl \{ (5 - 3 d) h^3 
\nonumber \\ &&
+ (22d - 38) h^2 Z + (44 - 12 d^2) h Z^2 
+ 8 (d-1)(5d-11) Z^3 \Bigr \} {\bf I}(h,Z,Z)
\nonumber \\ &&
+ \frac{g^2 (d-2)}{4 W (h - 4 W)} \bigl [ (6 - 6d) W^2 + (7 + d) h W - 2 h^2 \bigr ] {\bf A}(h) {\bf A}(W) 
\nonumber \\ &&
+ \frac{(g^2  + g^{\prime 2})(d-2)}{8 Z (h - 4 Z)} \bigl [ (6 - 6d) Z^2 + (7 + d) h Z - 2 h^2 \bigr ]
{\bf A}(h) {\bf A}(Z)
\nonumber \\ &&
+ 2 \lambda (d-1) (W+Z) \bigl [ 1 + (4d - 14) W/Z + W^2/Z^2 \bigr ] {\bf I}(0,W,Z)
\nonumber \\ &&
+ 2 \lambda (d-1) W \left[(3d-7) + (10 - 3d) W/Z - W^2/Z^2 \right ] {\bf I}(0,0,W)
\nonumber \\ &&
+ 2 \lambda (d-1) W (3 W/Z - 2) {\bf I}(0,0,Z)
+ \frac{g^2}{4 W Z^2} \Bigl \{
(4 - 2 d) Z^3 
+ 4 (d-2) (d-1) h W Z 
\nonumber \\ &&
+ (d-1) h (W-Z)^2  
+ 8 (d-2)^2 (d-1) W^2 Z 
- 8 (d-2)^2 W Z^2 \Bigr \} {\bf A}(W) {\bf A}(Z)
\nonumber \\ &&
+ \frac{(g^2 + g^{\prime 2})}{16 Z (h - 4 Z)}  \bigl [(3d - 5) h^2 + (28 - 16 d) h Z + 4 (d+3)(d-2) Z^2 \bigr ]
{\bf A}(Z)^2 
\nonumber \\ && 
+ \frac{g^2}{8 (d-3) W Z (h - 4 W)} \Bigl \{ (d-3)(3d-5) h^2 Z + 16 (d-2)^2 (d-1) h W^2 
\nonumber \\ && 
+ (8 d^4 - 88 d^3 + 300 d^2 - 384 d + 148) h W Z
+ 2 (d-3)(d-2) h Z^2
\nonumber \\ && 
- 64 (d-2)^2 (d-1) W^3 
- (d-2) (d-1) (32 d^2 - 260 d + 428) W^2 Z
\nonumber \\ && 
- 8 (d-3) (d-2) W Z^2 
\Bigr \} {\bf A}(W)^2 
\biggr ]
.
\label{eq:DeltaB2SM}
\eeq
For convenience, these expressions for $\Delta^{(1)}_B$ and $\Delta^{(2)}_B$ are also given in
the ancillary file {\tt SMDeltaGB.txt}.

For practical calculations, what is ultimately needed is the expansion of the $\Delta_B^{(n)}$ in $\epsilon$.
To obtain this, one first uses the counterterm relations eq.~(\ref{eq:XBrenorm}), expanding in $\kappa$ with the help of the derivatives of the master integrals with respect to their arguments, given in the ancillary file 
{\tt derivatives.txt} distributed with ref.~\cite{Martin:2016bgz}. Then, one applies the $\epsilon$ expansions of the master integrals using 
\beq
{\bf A}(x) &=& -\frac{x}{\epsilon} + A(x) + \epsilon A_{\epsilon}(x) + \epsilon^2 A_{\epsilon^2}(x) + \cdots
,
\label{eq:AAxexpeps}
\\
{\bf I}(x,y,z) &=& -\frac{1}{2\epsilon^2} (x + y + z) + \frac{1}{\epsilon} \bigl [A(x) + A(y) + A(z) - (x+y+z)/2 \bigr ]
\nonumber \\ &&
+ \bigl [ I(x,y,z) + A_\epsilon(x) + A_\epsilon(y) + A_\epsilon(z)\bigr ] + \epsilon I_\epsilon(x,y,z) + \cdots
,
\label{eq:IIxyzexpeps}
\eeq
and similar expansions for the three-loop master vacuum integrals given in ref.~\cite{Martin:2016bgz}. The results for the tadpole-free condition in bare perturbation theory can then be reorganized by collecting terms with the same powers of $\kappa$, giving the form
\beq
G_B = -\kappa \tilde \Delta^{(1)} -\kappa^2 \tilde \Delta^{(2)} -\kappa^3 \tilde \Delta^{(3)} + \cdots,
\label{eq:GBminexpanded}
\eeq
where terms at loop order $n$ have poles in $\epsilon$ up to order $n$, with expansions
\beq
\widetilde \Delta^{(1)} &=& 
\frac{1}{\epsilon} \widetilde \Delta^{(1,-1)} 
+ \widetilde \Delta^{(1,0)} 
+ \epsilon \widetilde \Delta^{(1,1)} 
+ \epsilon^2 \widetilde \Delta^{(1,2)} + \cdots 
,
\label{eq:DeltaB1SMexp}
\\
\widetilde\Delta^{(2)} &=& 
\frac{1}{\epsilon^2} \widetilde \Delta^{(2,-2)} 
+ \frac{1}{\epsilon} \widetilde \Delta^{(2,-1)} + \widetilde \Delta^{(2,0)} 
+ \epsilon \widetilde \Delta^{(2,1)} + \cdots
,
\label{eq:DeltaB2SMexp}
\\
\widetilde \Delta^{(3)} &=&
\frac{1}{\epsilon^3} \widetilde \Delta^{(3,-3)} 
+ \frac{1}{\epsilon^2} \widetilde \Delta^{(3,-2)} 
+ \frac{1}{\epsilon} \widetilde\Delta^{(3,-1)} + \widetilde\Delta^{(3,0)} + \cdots 
.
\label{eq:DeltaB3SMexp}
\eeq
The ellipses represent the terms higher order in $\epsilon$ that are needed if and only if one is going to four-loop order. For example, the one-loop contributions are
(now reverting back to the notation in which all parameters without $B$ subscripts are \MSbar renormalized, not bare,
with $t = y_t^2 \phi^2/2$ and $h = 2 \lambda \phi^2$ and $W = g^2 \phi^2/4$ and $Z = (g^2 + g^{\prime 2}) \phi^2/4$):
\beq
\widetilde \Delta^{(1,-1)} &=& 
2 N_c y_t^2 t - 3 \lambda h - \frac{3}{2} g^2 W - \frac{3}{4} (g^2 + g^{\prime 2}) Z  ,
\\
\widetilde \Delta^{(1,0)} &=& 
-2 N_c y_t^2 A(t) +  3 \lambda A(h) + \frac{g^2}{2} \left [3 A(W) + 2 W \right ]
+ \frac{g^2 + g^{\prime 2}}{4} \left [3 A(Z) + 2 Z \right ]
,
\\
\widetilde \Delta^{(1,1)} &=&
-2 N_c y_t^2 A_\epsilon(t) +  3 \lambda A_\epsilon(h) + \frac{g^2}{2} \left [3 A_\epsilon(W) - 2 A(W) \right ]
+ \frac{g^2 + g^{\prime 2}}{4} \left [3 A_\epsilon(Z) - 2 A(Z) \right ] 
,
\\
\widetilde \Delta^{(1,2)} &=& 
-2 N_c y_t^2 A_{\epsilon^2}(t) +  3 \lambda A_{\epsilon^2}(h) + \frac{g^2}{2} \left [3 A_{\epsilon^2}(W) - 2 A_{\epsilon}(W) \right ]
+ \frac{g^2 + g^{\prime 2}}{4} \left [3 A_{\epsilon^2}(Z) - 2 A_{\epsilon} (Z) \right ] 
.
\nonumber \\ &&
\eeq
The other $\widetilde\Delta^{(n,k)}$ with $n+k \leq 3$, which are the ones needed for a future calculation of the Higgs boson pole mass (and other observables) at three-loop order, are much more lengthy, and therefore are provided in an ancillary file {\tt SMGB.txt} distributed with this paper.

One can use the above results to recover the previously known minimization condition for the renormalized $G$.
Using eq.~(\ref{eq:XBrenorm}) with the coefficients found in the ancillary file {\tt SMcounters.txt}, one finds
that the \MSbar counterterm relationship between $G_B$ and $G$ is 
\beq
G_B &=& G + \frac{\kappa}{\epsilon} \bigl [ a_{11} + b_{11} G \bigr ]
+ \kappa^2 \Bigl (\frac{1}{\epsilon^2} \bigl [ a_{22} + b_{22} G \bigr ] 
+ \frac{1}{\epsilon} \bigl [ a_{21} + b_{21} G \bigr ]  \Bigr )
\nonumber \\ &&
+ \kappa^3 \Bigl (\frac{1}{\epsilon^3} \bigl [ a_{33} + b_{33} G \bigr ] 
+ \frac{1}{\epsilon^2} \bigl [ a_{32} + b_{32} G \bigr ]  
+ \frac{1}{\epsilon} \bigl [ a_{31} + b_{31} G \bigr ]  \Bigr )
+ {\cal O}(\kappa^4)
,
\label{eq:GBSMcounters}
\eeq
in which, for example, the one-loop coefficients are
\beq
a_{11} &=& 
-2 N_c y_t^2 t + 3 \lambda h + \frac{3}{4} (g^2 + g^{\prime 2}) Z + \frac{3}{2} g^2 W ,
\\
b_{11} &=& N_c y_t^2 + 6 \lambda - \frac{3}{4} (g^2 + g^{\prime 2})  - \frac{3}{2} g^2 .
\eeq
The complete results for eq.~(\ref{eq:GBSMcounters}) are provided in the ancillary file {\tt SMcounters.txt}.
One can now solve eq.~(\ref{eq:GBSMcounters}) iteratively for $G$ in the form of eq.~(\ref{eq:GminDelta}).
An important consistency check is that the quantities $\Delta^{(n)}$ are free of poles in $\epsilon$; this
relies on the following equalities:
\beq
\widetilde\Delta^{(1,-1)} &=& -a_{11},
\\
\widetilde\Delta^{(2,-2)} &=& -a_{22},
\\
\widetilde\Delta^{(2,-1)} &=& -a_{21} 
+ b_{11} \widetilde \Delta^{(1,0)},
\\
\widetilde\Delta^{(3,-3)} &=& -a_{33},
\\
\widetilde\Delta^{(3,-2)} &=& -a_{32} 
+ b_{22} \widetilde \Delta^{(1,0)},
\\
\widetilde\Delta^{(3,-1)} &=& -a_{31} 
+ b_{11} \widetilde \Delta^{(2,0)}
+ b_{21} \widetilde \Delta^{(1,0)}
+ \left [b_{22} - (b_{11})^2 \right ] \widetilde \Delta^{(1,1)}.
\eeq
Then one finds that the quantities appearing in eq.~(\ref{eq:GminDelta}) are
\beq
\Delta^{(1)} &=& \widetilde \Delta^{(1,0)} + {\cal O}(\epsilon)
,
\\
\Delta^{(2)} &=& \widetilde \Delta^{(2,0)} - b_{11} \widetilde \Delta^{(1,1)} + {\cal O}(\epsilon)
,
\\
\Delta^{(3)} &=& \widetilde \Delta^{(3,0)} 
- b_{11} \widetilde \Delta^{(2,1)}
- b_{21} \widetilde \Delta^{(1,1)} 
- \left [b_{22} - (b_{11})^2 \right ] \widetilde \Delta^{(1,2)} + {\cal O}(\epsilon)
,
\label{eq:Delta3fromtildes}
\eeq
which, in the limit $\epsilon \rightarrow 0$, are the same as already given in ref.~\cite{Martin:2017lqn},
and in an independent way through two-loop order in ref.~\cite{Espinosa:2017aew}.

However, as noted in the Introduction, if one is working in the bare scheme to calculate some other quantity, one should use the minimization condition for $G_B$, which in expanded form is provided for the Standard Model by eqs.~(\ref{eq:GBminexpanded})-(\ref{eq:DeltaB3SMexp}) above with explicit results in the ancillary file {\tt SMGB.txt}. In that sense, the expressions for $\widetilde \Delta^{(n,k)}$ provided with the present paper are more complete and useful information than the renormalized $\Delta^{(n)}$ given in ref.~\cite{Martin:2017lqn}. For example, the terms $\widetilde \Delta^{(1,1)}$, $\widetilde \Delta^{(1,2)}$, and $\widetilde \Delta^{(2,1)}$ contribute nontrivially to the calculation of the complete three-loop pole mass of the Higgs boson, to be reported in a future paper. For future calculations of the complete three-loop pole masses of the $W$ and $Z$ bosons and the top quark, of the terms with positive powers of $\epsilon$ in eqs.~(\ref{eq:DeltaB1SMexp})-(\ref{eq:DeltaB3SMexp}), only $\widetilde \Delta^{(1,1)}$ is needed, because $G_B$ does not appear in the tree-level part in each of those cases. 

It is useful to note that $\Delta^{(2)}$ does not contain any functions $A_\epsilon(x)$,
even though $\widetilde \Delta^{(2,0)}$ and $\widetilde \Delta^{(1,1)}$ each do.
Similarly, all instances of $A_\epsilon(x)$, $A_{\epsilon^2}(x)$, and $I_{\epsilon}(x,y,z)$
cancel from $\Delta^{(3)}$, even though they are present in the individual contributions on the right side of eq.~(\ref{eq:Delta3fromtildes}). As observed e.g.~in refs.~\cite{Martin:2001vx,Martin:2017lqn,Martin:2021pnd}, this cancellation is a general feature, directly related to the cancellation of poles in $\epsilon$ and the particular subtractions of subdivergences used to define the renormalized loop integrals $A(x)$ and $I(x,y,z)$. The eventual cancellation of $A_\epsilon(x)$, $A_{\epsilon^2}(x)$, and $I_{\epsilon}(x,y,z)$ in three-loop renormalized quantities should provide a useful practical check.  

%%\color{blue}%%NEW
Although the applications motivating this paper center around the minimization condition for the bare effective potential, it is also interesting to consider the form of the renormalized effective potential following from the Goldstone boson anti-resummation prescription. The renormalized effective potential (as opposed to its renormalized minimization condition, as just discussed) does contain poles in $\epsilon$, corresponding to IR divergences. This should not be surprising or worrisome, because the effective potential itself is not an observable (recall that it is not even gauge invariant) and is therefore not required to be infrared safe. For example, applying the \MSbar parameter redefinitions through one-loop order to eqs.~(\ref{eq:Vbare0SM}) and (\ref{eq:V1SMpreHexp}), one readily finds
\beq
V^{(1)} &=&  
\frac{3}{4} Z^2 \Bigl (\lnbar(Z)  - 5/6 \Bigr )
+
\frac{3}{2} W^2 \Bigl (\lnbar(W)  - 5/6 \Bigr )   
-N_c\hspace{1pt} t^2 \Bigl (\lnbar(t)  - 3/2 \Bigr ) 
\nonumber \\ &&
+ \frac{1}{4} (h + G)^2 \Bigl (\lnbar(h+G)  - 3/2 \Bigr )
+ \frac{3}{4 \epsilon_{\rm IR}} G^2.
\eeq
The last term is an IR divergence, associated with the use of massless propagators for the Goldstone fields in the anti-resummation prescription. It corresponds to the replacement
\beq
\frac{3}{4} G^2 \Bigl (\lnbar(G)  - 3/2 \Bigr ) &\rightarrow& \frac{3}{4 \epsilon_{\rm IR}} G^2
\eeq
in the usual renormalized one-loop effective potential computed without any resummation or anti-resummation. This can be traced to the fact that the vanishing dimensional regularization integral of $1/(p^2)^2$ appearing in eq.~(\ref{eq:AGBantiresummation}) can be split into equal and opposite UV and IR divergences,
\beq
\int_p \frac{1}{(p^2)^2} &=& \frac{1}{\epsilon_{\rm UV}} - \frac{1}{\epsilon_{\rm IR}} .
\eeq
Thus, with the anti-resummation prescription, \MSbar renormalization eliminates the $1/\epsilon_{\rm UV}$ poles as usual, leaving behind the $1/\epsilon_{\rm IR}$ pole as a sort of proxy for the $\lnbar(G)$ singularity found in the usual approach.

In a similar way, one finds that the renormalized 2-loop effective potential with the anti-resummation prescription
has the following infrared pole structure
\beq
V^{(2)}_{\rm poles} &=&
\frac{G}{\epsilon} P_{11}^{(2)} + G^2 \left ( \frac{1}{\epsilon^2} P_{22}^{(2)} + \frac{1}{\epsilon} P_{21}^{(2)} \right )
+ {\cal O}(G^3),
\eeq
with
\beq
P_{11}^{(2)} &=& 
\frac{9}{2} \lambda A(h) -3 N_c y_t^2 A(t) 
+ \frac{9}{8} (g^2 + g^{\prime 2}) \left [A(Z) + 2 Z/3 \right ]
+ \frac{9}{4} g^2 \left [A(W) + 2 W/3 \right]
,
\\
P_{22}^{(2)} &=&
\frac{9}{4} \lambda - \frac{9g^2 g^{\prime 2}}{8(g^2 + g^{\prime 2})}
,
\\
P_{21}^{(2)} &=&
3 (g^2 + g^{\prime 2}) \frac{A(Z) - A(W)}{Z - W}  
- \frac{3}{8} (g^2 + g^{\prime 2}) \frac{A(h) - A(Z)}{h-Z}
- \frac{3}{4} g^2 \frac{A(h) - A(W)}{h-W} 
\nonumber \\ &&
+ N_c y_t^2 \left [1 + 3 A(t)/2 t \right ] 
+ 3 \lambda \left [1 + 3 A(h)/2 h \right ]
+ 3 g^2 A(W)/2W
\nonumber \\ &&
- \frac{3 (9 g^4 + 11 g^2 g^{\prime 2} + 5 g^{\prime 4})}{4(g^2 + g^{\prime 2})} \frac{A(Z)}{Z} 
- \frac{3 (19 g^4 + 8 g^2 g^{\prime 2} + g^{\prime 4})}{16 (g^2 + g^{\prime 2})}
.
\eeq
Despite these poles in the renormalized effective potential in the anti-resummation prescription, the corresponding minimization condition for the renormalized $G$ is free of poles as $\epsilon \rightarrow 0$,
because 
\beq
0 &=& \frac{1}{\phi}\frac{\partial}{\partial \phi} \left ( V^{(0)} + \kappa V^{(1)} + \kappa^2 V^{(2)} + \ldots \right ) 
\eeq
can be written in a factorized form,
\beq
0 &=& 
\Bigl \{ 
G + \kappa \Delta^{(1)} + \kappa^2 \Delta^{(2)} + \ldots 
\Bigr \}
\Bigl \{
1 + \frac{\kappa}{\epsilon} 3 \lambda + \kappa^2 \Bigl (\frac{1}{\epsilon^2} Q^{(2)}_{22} + \frac{1}{\epsilon} Q^{(2)}_{21} \Bigr )
 + \ldots 
\Bigr \}
,
\eeq
up to terms of higher order in $\kappa$ and $G$, where
\beq
Q^{(2)}_{22} &=& 4 \lambda P_{22}^{(2)},
\\
Q^{(2)}_{21} &=& 4 \lambda P_{21}^{(2)} + \frac{1}{\phi}\frac{\partial}{\partial \phi} P_{11}^{(2)}.
\eeq
However, I reiterate that for the applications that constitute the motivation for the present paper, the bare effective potential, and specifically its minimization condition,
is the more directly useful construction, and that the important thing is that the IR divergences cancel completely (along with UV divergences) from renormalized expressions for observables such as pole masses.
%%\color{black}%%ENDNEW

\section{Goldstone boson anti-resummation in minimal supersymmetry\label{subsec:MSSM}}
\setcounter{equation}{0}
\setcounter{figure}{0}
\setcounter{table}{0} 
\setcounter{footnote}{1}

The most general use of the effective potential can have some complications that do not occur in the case of the Standard Model discussed in the previous section. There can be more than one vacuum expectation value minimization condition, and more than one distinct Goldstone-boson squared mass. 
To illustrate these complications and how they are dealt with, I will now discuss how Goldstone boson anti-resummation works in a case with two distinct fields with vacuum expectation values, the minimal supersymmetric Standard Model (MSSM). For a review of the MSSM with notation consistent with the present paper, see ref.~\cite{Martin:1997ns}. The explicit result for the MSSM two-loop effective potential was given in ref.~\cite{Martin:2002iu}, and its Goldstone boson resummation was discussed in the old way in \cite{Kumar:2016ltb}. The generalization at two-loop order to arbitrary theories has been given in ref.~\cite{Braathen:2016cqe}. Method (iii) in section 3.1.3 of that paper seems to be somewhat similar in spirit to the approach proposed in the present paper, but with a rather different implementation in which logarithms of the Goldstone-boson squared mass still appear in intermediate steps.

For typographical simplicity, all of the Lagrangian quantities (and quantities directly related to them) throughout this section are bare parameters, with the subscript $B$ omitted to reduce clutter.

The MSSM contains two Higgs doublet fields, whose neutral component background fields (equal to the VEVs, at the minimum of the effective potential) will be denoted $v_u$ and $v_d$. The tree-level bare potential for $v_u$ and $v_d$ is
\beq
V_B^{(0)} &=& (|\mu|^2 + m^2_{H_u}) v_u^2 + (|\mu|^2 + m^2_{H_d}) v_d^2 - 2 b v_u v_d + 
\frac{1}{8} (g^2 + g^{\prime 2}) (v_u^2 - v_d^2)^2
,
\label{eq:VBtreeMSSM}
\eeq
where $\mu$ is the superpotential Higgs mass, $b$ is a supersymmetry-breaking holomorphic Higgs squared mass,
and $m^2_{H_u}, m^2_{H_d}$ are non-holomorphic squared masses, and $g$, $g'$ are the $SU(2)_L$ and $U(1)_Y$
gauge couplings. One also writes
\beq
s_\beta \,=\, {v_u}/{\sqrt{v_u^2 + v_d^2}} ,
\qquad\quad
c_\beta \,=\, {v_d}/{\sqrt{v_u^2 + v_d^2}} ,
\eeq
defining an angle $\beta$ for which $s_\beta$ and $c_\beta$ are the sine and cosine. The normalization of $v_u$ and $v_d$ is such
that the $W$ and $Z$ boson bare squared masses are
\beq
W &=& \frac{1}{2} g^2 (v_u^2 + v_d^2),
\\
Z &=& \frac{1}{2} (g^2 + g^{\prime 2}) (v_u^2 + v_d^2).
\eeq

In the presence of the background fields $v_u$ and $v_d$, the complex doublets $H_u = (H_u^+, H_u^0)$
and $H_d = (H_d^0, H_d^-)$ mix to form squared mass eigenstates $G^0, G^\pm$ (the Goldstone bosons), $h$ and $H$ (the neutral scalar Higgs bosons), $A$ (the neutral pseudoscalar Higgs boson), and $H^\pm$ (the charged Higgs bosons), with
\beq
\begin{pmatrix}
H_u^0\\
H_d^0
\end{pmatrix}
&=&
\begin{pmatrix}
v_u
\\
v_d
\end{pmatrix}
+ \frac{1}{\sqrt{2}} R_{\alpha_0} 
\begin{pmatrix}
h^0\\
H^0
\end{pmatrix}
+ \frac{i}{\sqrt{2}} R_{\beta_0} 
\begin{pmatrix}
G^0\\
A^0
\end{pmatrix},
\\
\begin{pmatrix}
H_u^+\\
H_d^{-*}
\end{pmatrix}
&=&
R_{\beta_\pm}
\begin{pmatrix}
G^+\\
H^+
\end{pmatrix},
\eeq
in terms of orthogonal mixing matrices
\beq
R_{\beta_0} &=& 
\begin{pmatrix} 
s_{\beta_0} & c_{\beta_0}
\\
-c_{\beta_0} & s_{\beta_0}
\end{pmatrix}
,
\qquad\>\>\>
R_{\beta_\pm} \>=\> 
\begin{pmatrix} 
s_{\beta_\pm} & c_{\beta_\pm}
\\
-c_{\beta_\pm} & s_{\beta_\pm}
\end{pmatrix}
,
\qquad\>\>\>
R_{\alpha_0} \>=\> 
\begin{pmatrix} 
c_{\alpha_0} & s_{\alpha_0}
\\
-s_{\alpha_0} & c_{\alpha_0}
\end{pmatrix}
,
\label{eq:Rbeta0RbetapmRalpha0}
\eeq
where explicit formulas for $\beta_0$, $\beta_\pm$, and $\alpha_0$ will be given below. (Note that $\beta_0$ and $\beta_\pm$ differ from $\beta$ by loop-suppressed amounts, because the VEVs $v_u$ and $v_d$ are not the minimum of the tree-level bare potential.)
With the usual organization of bare perturbation theory, the propagators in the Higgs field sector have squared masses
\beq
m^2_{G^0} &=& 
|\mu|^2 + \frac{1}{2} (m_{H_u}^2 + m_{H_d}^2) 
- \frac{1}{2} \Big \{ \bigl [m_{H_u}^2 - m_{H_d}^2 + 
\frac{1}{2} (g^2 + g^{\prime 2}) (v_u^2 - v_d^2) \bigr ]^2 + 4 b^2\Big\}^{1/2} 
,
\label{eq:m2G0}
\\
m^2_{G^\pm} &=& 
|\mu|^2 + \frac{1}{2} (m_{H_u}^2 + m_{H_d}^2) + 
\frac{1}{4} g^2 (v_u^2 + v_d^2)
\nonumber \\ &&
- \frac{1}{2} \Big\{ \bigl [m_{H_u}^2 - m_{H_d}^2 + \frac{1}{2} 
g^{\prime 2}
(v_u^2 - v_d^2) \bigr ]^2 + (2b+ g^2 v_u v_d)^2\Big\}^{1/2} 
,
\label{eq:m2Gp}
\\
m^2_{h,H} &=& 
|\mu|^2 + \frac{1}{2} (m_{H_u}^2 + m_{H_d}^2) + \frac{1}{4} (g^2 + g^{\prime 2}) (v_u^2 + v_d^2)
\nonumber \\ &&
\mp \frac{1}{2} \Big\{ \bigl [
m_{H_u}^2 - m_{H_d}^2 + (g^2 + g^{\prime 2}) (v_u^2 - v_d^2)\bigr ]^2 
+(2b + (g^2 + g^{\prime 2}) v_u v_d )^2\Big \}^{1/2}
,
\\
m^2_{A} &=& 
|\mu|^2 + \frac{1}{2} (m_{H_u}^2 + m_{H_d}^2)
+ \frac{1}{2} \Big\{ \bigl [m_{H_u}^2 - m_{H_d}^2 + 
\frac{1}{2} (g^2 + g^{\prime 2}) (v_u^2 - v_d^2) \bigr ]^2 + 4 b^2 \Big \}^{1/2}
,
\phantom{xxx}
\\
m^2_{H^\pm} &=& 
|\mu|^2 + \frac{1}{2} (m_{H_u}^2 + m_{H_d}^2) + 
\frac{1}{4} g^2 (v_u^2 + v_d^2)
\nonumber \\ &&
+ \frac{1}{2} \Big\{ \bigl [m_{H_u}^2 - m_{H_d}^2 + 
\frac{1}{2} g^{\prime 2} (v_u^2 - v_d^2) \bigr ]^2 + (2b+ g^2 v_u v_d)^2\Big \}^{1/2}
.
\eeq
Note that unlike the case in the Standard Model where the charged and neutral Goldstone bosons arise from a single gauge multiplet, in the MSSM (and more general two-Higgs-doublet models), the charged and neutral Goldstone bosons have different squared masses. 

Now define the quantities
\beq
G_u &\equiv& 
\frac{1}{2 v_u} \frac{\partial}{\partial v_u} V_B^{(0)} \,=\, 
|\mu|^2 + m^2_{H_u} - b \frac{v_d}{v_u} + \frac{1}{4} (g^2 + g^{\prime 2}) (v_u^2 - v_d^2),
\label{eq:Gu}
\\
G_d &\equiv& 
\frac{1}{2 v_d} \frac{\partial}{\partial v_d} V_B^{(0)} \,=\,
|\mu|^2 + m^2_{H_d} - b \frac{v_u}{v_d} - \frac{1}{4} (g^2 + g^{\prime 2}) (v_u^2 - v_d^2).
\label{eq:Gd}
\eeq
Using these to eliminate $m^2_{H_u}$ and $m^2_{H_d}$ from eqs.~(\ref{eq:m2G0}) and (\ref{eq:m2Gp}) and expanding 
in small $G_u$ and $G_d$ (since they vanish at the minimum of the potential in the tree-level approximation),
one imposes the rules for Goldstone boson anti-resummation:
\beq
\frac{1}{p^2 + m^2_{G^0}} &\rightarrow& \frac{1}{p^2} - \frac{G^0}{(p^2)^2} + \frac{(G^0)^2}{(p^2)^3} + \cdots
,
\label{eq:expandG0prop}
\\
\frac{1}{p^2 + m^2_{G^\pm}} &\rightarrow& \frac{1}{p^2} - \frac{G^\pm}{(p^2)^2} + \frac{(G^\pm)^2}{(p^2)^3} + \cdots
,
\label{eq:expandGpmprop}
\eeq
where
\beq
G^0 &=& \left (s_\beta^2 G_u + c_\beta^2 G_d \right ) - \frac{s_{\beta}^2 c_\beta^2}{A} (G_u - G_d)^2 
+ \frac{s_\beta^2 c_\beta^2 (c_\beta^2 - s_\beta^2)}{A^2} (G_u - G_d)^3
+
\ldots
,
\\
G^\pm &=& \left (s_\beta^2 G_u + c_\beta^2 G_d \right ) - \frac{s_{\beta}^2 c_\beta^2}{H^\pm} (G_u - G_d)^2 
+ \frac{s_\beta^2 c_\beta^2 (c_\beta^2 - s_\beta^2)}{(H^\pm)^2} (G_u - G_d)^3
+
\ldots
,
\eeq
in which we have defined
\beq
A \,=\, b/s_\beta c_\beta,
\qquad\quad
H^\pm \,=\, A + W.
\label{eq:defAHpm}
\eeq
Thus, one adopts a version of bare perturbation theory in which the Goldstone bosons have propagator
masses that are exactly 0, and have two-point interactions that involve the loop-suppressed quantities
$G_u$ and $G_d$. Unlike the case of the Standard Model, the terms linear in $G^0$ and $G^\pm$ in the replacement rules of eqs.~(\ref{eq:expandG0prop}) and (\ref{eq:expandGpmprop}) are nonlinear combinations of $G_u$ and $G_d$, but the relation is still expressed using only non-negative integer powers. More generally, in theories with multiple Goldstone fields and VEVs, the Goldstone-boson squared masses can be expanded in terms of the tree-level tadpole quantities $G_n$, as shown e.g.~in ref.~\cite{Braathen:2016cqe}.

It is also convenient to deconstruct the non-Goldstone Higgs scalar propagators, in a way analogous to 
eq.~(\ref{eq:HBhBexpansion}) in the Standard Model. This is done according to
\beq
\frac{1}{p^2 + m^2_{A^0}} &\rightarrow& \frac{1}{p^2 + A} 
+ \frac{\delta_A}{(p^2 + A)^2} 
+ \frac{\delta_A^2}{(p^2 + A)^3} + \cdots,
\\
\frac{1}{p^2 + m^2_{H^\pm}} &\rightarrow& \frac{1}{p^2 + H^\pm} 
+ \frac{\delta_{H^\pm}}{(p^2 + H^\pm)^2} 
+ \frac{\delta_{H^\pm}^2}{(p^2 + H^\pm)^3} + \cdots,
\\
\frac{1}{p^2 + m^2_{h}} &\rightarrow& \frac{1}{p^2 + h} 
+ \frac{\delta_h}{(p^2 + h)^2} 
+ \frac{\delta_h^2}{(p^2 + h)^3} + \cdots,
\\
\frac{1}{p^2 + m^2_{H}} &\rightarrow& \frac{1}{p^2 + H} 
+ \frac{\delta_H}{(p^2 + H)^2} 
+ \frac{\delta_H^2}{(p^2 + H)^3} + \cdots,
\eeq
where $A$ and $H^\pm$ were given in eq.~(\ref{eq:defAHpm}), and
\beq
h,H &=& \frac{1}{2} \left [A + Z \mp \sqrt{(A+Z)^2 - 4 (s_\beta^2 - c_\beta^2)^2 A Z}  \right ]
,
\label{eq:defhH}
\eeq
and
\beq
\delta_A &=& -\left (c_\beta^2 G_u + s_\beta^2 G_d \right ) 
- \frac{s_{\beta}^2 c_\beta^2}{A} (G_u - G_d)^2 + \cdots
,
\\
\delta_{H^\pm} &=& -\left (c_\beta^2 G_u + s_\beta^2 G_d \right ) 
- \frac{s_{\beta}^2 c_\beta^2}{H^\pm} (G_u - G_d)^2 + \cdots
,
\\
\delta_{h} &=& -\frac{1}{2} (G_u + G_d) + (c_\beta^2 - s_\beta^2)\frac{A-Z}{2(H-h)} (G_u - G_d) 
+ s_\beta^2 c_\beta^2 \frac{(A+Z)^2}{(H-h)^3} (G_u - G_d)^2  + \cdots
,
\\
\delta_{H} &=& -\frac{1}{2} (G_u + G_d) + (s_\beta^2 - c_\beta^2)\frac{A-Z}{2(H-h)} (G_u - G_d)
- s_\beta^2 c_\beta^2 \frac{(A+Z)^2}{(H-h)^3} (G_u - G_d)^2  + \cdots
.
\phantom{XXX}
\eeq
The mixing angles of the Goldstone and Higgs scalars are given exactly by
\beq
\cot(2 \beta_0) &=& \cot(2 \beta) + \frac{G_u- G_d}{2 b},
\\
\cot(2 \beta_\pm) &=& \cot(2 \beta) + \frac{G_u- G_d}{2 b + g^2 v_u v_d},
\\
\cot(2 \alpha_0) &=& \cot(2 \alpha) + \frac{G_u- G_d}{2 b + (g^2 + g^{\prime 2}) v_u v_d},
\eeq
where
\beq
\cot(2\alpha) = \left (\frac{A-Z}{A+Z} \right ) \cot(2\beta).
\eeq
The sines and cosines of the mixing angles $\beta_0$, $\beta_\pm$, and $\alpha_0$ in the matrices in eq.~(\ref{eq:Rbeta0RbetapmRalpha0}), which appear in three-point and four-point interaction couplings of the Goldstone and Higgs mass eigenstates, can now each be expanded in a power series in $G_u$ and $G_d$, according to
\beq
s_{\beta_0} &=& s_\beta \Bigl [ 1 - c_\beta^2 \left ( \frac{G_u - G_d}{A} \right )
+ c_\beta^2 (1 - 5 s_\beta^2/2) \left ( \frac{G_u - G_d}{A} \right )^2
+ \ldots
\Bigr ]
,
\\
c_{\beta_0} &=& c_\beta \Bigl [ 1 + s_\beta^2 \left (\frac{G_u - G_d}{A}\right )
+ s_\beta^2 (1 - 5 c_\beta^2/2) \left (\frac{G_u - G_d}{A} \right )^2
+ \ldots
\Bigr ]
,
\\
s_{\beta_\pm} &=& s_\beta \Bigl [ 1 - c_\beta^2 \left (\frac{G_u - G_d}{A+W} \right )
+ c_\beta^2 (1 - 5 s_\beta^2/2) \left ( \frac{G_u - G_d}{A+W} \right )^2
+ \ldots
\Bigr ]
,
\\
c_{\beta_\pm} &=& c_\beta \Bigl [ 1 + s_\beta^2 \left (\frac{G_u - G_d}{A+W} \right )
+ s_\beta^2 (1 - 5 c_\beta^2/2) \left ( \frac{G_u - G_d}{A+W} \right )^2
+ \ldots
\Bigr ]
,
\\
s_{\alpha_0} &=& s_\alpha \Bigl [ 1 - 
\frac{s_\alpha c_\alpha^3}{s_\beta c_\beta} \left ( \frac{G_u - G_d}{A+Z} \right )
+ \frac{s_\alpha^2 c_\alpha^4 (1 - 5 s_\alpha^2/2)}{s^2_\beta c^2_\beta} \left ( \frac{G_u - G_d}{A+Z} \right )^2
+ \ldots
\Bigr ]
,
\\
c_{\alpha_0} &=& c_\alpha \Bigl [ 1 + \frac{s_\alpha^3 c_\alpha}{s_\beta c_\beta} \left (\frac{G_u - G_d}{A+Z} \right )
+ \frac{s_\alpha^4 c_\alpha^2 (1 - 5 c_\alpha^2/2)}{s_\beta^2 c_\beta^2} \left ( \frac{G_u - G_d}{A+Z} \right )^2
+ \ldots
\Bigr ]
.
\label{eq:calpha0exp}
\eeq

To summarize, to implement the anti-resummation prescription, the bare propagators of the Goldstone bosons in the MSSM always have 0 squared masses, while those of the Higgs bosons have squared masses $A$, $H^\pm$, $h$, and $H$ given by eqs.~(\ref{eq:defAHpm}) and (\ref{eq:defhH}). The remaining portions of the Goldstone and Higgs boson squared masses are treated as two-point interactions, which gives results that are power series in the loop-suppressed quantities $G_u$ and $G_d$, defined as derivatives of the tree-level bare potential according to eqs.~(\ref{eq:Gu}) and (\ref{eq:Gd}). The other Higgs and Goldstone interactions are also power series in $G_u$ and $G_d$. As a result, the full effective potential can be written in the form of a series with only non-negative integer powers of the tree-level tadpole quantities $G_u$ and $G_d$, 
\beq
V_{\rm eff} &=& V_B^{(0)} + \sum_{n=1}^\infty \kappa^n \sum_{j=0}^\infty \sum_{k=0}^\infty
G_u^j \hspace{1pt} G_d^k \hspace{1pt} V_B^{(n,j,k)},
\eeq
where the $V_B^{(n,j,k)}$ do not depend on $m_{H_u}^2$ and $m^2_{H_d}$ (or equivalently $G_u$ and $G_d$, or $G^0$ and $G^\pm$). The minimization conditions for the full effective potential in the bare perturbation theory are therefore
\beq
G_u &=& -\sum_{n=1}^\infty \kappa^n \sum_{j=0}^\infty \sum_{k=0}^\infty G_u^j\hspace{1pt} G_d^k\hspace{1pt} \delta_u^{(n,j,k)},
\label{eq:Gudelta}
\\
G_d &=& -\sum_{n=1}^\infty \kappa^n \sum_{j=0}^\infty \sum_{k=0}^\infty G_u^j \hspace{1pt} G_d^k \hspace{1pt}\delta_d^{(n,j,k)},
\label{eq:Gddelta}
\eeq
where
\beq
\delta_u^{(n,j,k)} &=& \frac{1}{2 v_u} 
\left ( \frac{\partial V_B^{(n,j,k)}}{\partial v_u} + (j+1) V_B^{(n,j+1,k)} \frac{\partial G_u}{\partial v_u}
+ (k+1) V_B^{(n,j,k+1)} \frac{\partial G_d}{\partial v_u} \right )
,
\\
\delta_d^{(n,j,k)} &=& \frac{1}{2 v_d} 
\left ( \frac{\partial V_B^{(n,j,k)}}{\partial v_d} + (j+1) V_B^{(n,j+1,k)} \frac{\partial G_u}{\partial v_d}
+ (k+1) V_B^{(n,j,k+1)} \frac{\partial G_d}{\partial v_d} \right ),
\eeq
in which
\beq
\frac{1}{2 v_u} \frac{\partial G_u}{\partial v_u} &=& \frac{v_d b}{2 v_u^3} + \frac{1}{4} (g^2 + g^{\prime 2}),
\\
\frac{1}{2 v_d} \frac{\partial G_d}{\partial v_d} &=& \frac{v_u b}{2 v_d^3} + \frac{1}{4} (g^2 + g^{\prime 2}),
\\
\frac{1}{2 v_u} \frac{\partial G_d}{\partial v_u} &=& 
\frac{1}{2 v_d} \frac{\partial G_u}{\partial v_d} \>=\>
-\frac{b}{2 v_u v_d} - \frac{1}{4} (g^2 + g^{\prime 2})
.
\eeq
Equations (\ref{eq:Gudelta}) and (\ref{eq:Gddelta}) can now easily be solved, by substitution and matching powers of $\kappa$, in the form
\beq
G_u &=& -\sum_{n=1}^\infty \kappa^n \Delta^{(n)}_u,
\label{eq:GuDelta}
\\
G_d &=& -\sum_{n=1}^\infty \kappa^n \Delta^{(n)}_d,
\label{eq:GdDelta}
\eeq
where $\Delta^{(n)}_u$ and $\Delta^{(n)}_d$ do not depend on $G_u$ and $G_d$ (or $m^2_{H_u}$ and $m^2_{H_d}$,
or $G^0$ and $G^\pm$) at all. At three-loop order, the solution is
\beq
\Delta_u^{(1)} &=& \delta_u^{(1,0,0)},
\\
\Delta_u^{(2)} &=& \delta_u^{(2,0,0)} - \delta_u^{(1,0,0)} \delta_u^{(1,1,0)} - \delta_u^{(1,0,1)} \delta_d^{(1,0,0)},
\\
\Delta_u^{(3)} &=& \delta_u^{(3,0,0)} 
- \delta_u^{(1,0,0)} \delta_u^{(2,1,0)}
- \delta_u^{(1,1,0)} \delta_u^{(2,0,0)} 
- \delta_d^{(1,0,0)} \delta_u^{(2,0,1)}
- \delta_d^{(2,0,0)} \delta_u^{(1,0,1)}
\nonumber \\ &&
+ \delta_d^{(1, 0, 0)} \delta_d^{(1, 0, 1)} \delta_u^{(1, 0, 1)}
+ \delta_d^{(1, 1, 0)} \delta_u^{(1, 0, 0)} \delta_u^{(1, 0, 1)} 
+ \delta_d^{(1, 0, 0)} \delta_u^{(1, 0, 1)} \delta_u^{(1, 1, 0)}
\nonumber \\ &&
+ \delta_d^{(1, 0, 0)} \delta_u^{(1, 0, 0)} \delta_u^{(1, 1, 1)}
+ \delta_u^{(1,2,0)} \left [\delta_u^{(1,0,0)}\right]^2 
+ \delta_u^{(1,0,2)} \left [\delta_d^{(1,0,0)}\right]^2 
+ \delta_u^{(1,0,0)} \left [\delta_u^{(1,1,0)}\right]^2 ,
\phantom{xxx}
\eeq
and
\beq
\Delta_d^{(1)} &=& \delta_d^{(1,0,0)},
\\
\Delta_d^{(2)} &=& \delta_d^{(2,0,0)} - \delta_d^{(1,0,0)} \delta_d^{(1,0,1)} - \delta_d^{(1,1,0)}\delta_u^{(1,0,0)},
\\
\Delta_d^{(3)} &=& \delta_d^{(3,0,0)} 
- \delta_d^{(1,0,0)} \delta_d^{(2,0,1)}
- \delta_d^{(1,0,1)} \delta_d^{(2,0,0)} 
- \delta_u^{(1,0,0)} \delta_d^{(2,1,0)}
- \delta_u^{(2,0,0)} \delta_d^{(1,1,0)}
\nonumber \\ &&
+ \delta_u^{(1, 0, 0)} \delta_u^{(1, 1, 0)} \delta_d^{(1, 1, 0)}
+ \delta_u^{(1, 0, 1)} \delta_d^{(1, 0, 0)} \delta_d^{(1, 1, 0)} 
+ \delta_u^{(1, 0, 0)} \delta_d^{(1, 1, 0)} \delta_d^{(1, 0, 1)}
\nonumber \\ &&
+ \delta_u^{(1, 0, 0)} \delta_d^{(1, 0, 0)} \delta_d^{(1, 1, 1)}
+ \delta_d^{(1,0,2)} \left [\delta_d^{(1,0,0)}\right]^2 
+ \delta_d^{(1,2,0)} \left [\delta_u^{(1,0,0)}\right]^2 
+ \delta_d^{(1,0,0)} \left [\delta_d^{(1,0,1)}\right]^2 .
\phantom{xxx}
\eeq

As a reminder, all of the quantities appearing above in this section are bare ones, with the subscript $B$ omitted. Although explicit results for the MSSM will not be given here, the procedure should be clear. After computing $V_B^{(n)}$ using dimensional reduction, employing the Goldstone rules of section \ref{sec:Goldstonerules} to the
basis integrals to get the $V_B^{(n,j,k)}$ and thus $\delta_u^{(n,j,k)}$ and $\delta_d^{(n,j,k)}$, one expresses the results for $\Delta_u^{(n)}$ and $\Delta_d^{(n)}$ in terms of renormalized Lagrangian parameters and VEVs by substituting in the parameter redefinitions in the form eq.~(\ref{eq:XBrenorm}). This process involves evaluating the resulting derivatives of the basis integrals using the formulas in the ancillary file {\tt derivatives.txt} of ref.~\cite{Martin:2016bgz}. Finally, the bare master integrals ${\bf A}(x)$, ${\bf I}(x,y,z)$, ${\bf F}(w,x,y,z)$, ${\bf G}(v,w,x,y,z)$,
${\bf H}(u,v,w,x,y,z)$,  ${\bf H}'(u,v,w,x,y,z)$ are expanded in $\epsilon$, using the
expansions specified in ref.~\cite{Martin:2016bgz}, except for the special cases provided in section \ref{sec:special} above. The results contain renormalized master integrals $A(x)$, $I(x,y,z)$, $F(w,x,y,z)$, $\overline F(0,x,y,z)$, $G(v,w,x,y,z)$, $H(u,v,w,x,y,z)$ which can be evaluated numerically using the public code {\tt 3VIL}, also discussed in ref.~\cite{Martin:2016bgz}. Also appearing in the expansion are integral functions $A_{\epsilon}(x)$, $A_{\epsilon^2}(x)$, and $I_{\epsilon}(x,y,z)$ from eqs.~(\ref{eq:AAxexpeps}) and (\ref{eq:IIxyzexpeps}), but those always cancel from the final results for observables, and therefore do not need numerical evaluation. This process avoids all Goldstone singularities and spurious imaginary parts, as there are never any non-positive powers of $G_u$ and $G_d$ (or $G^0$ and $G^\pm$) at any stage. Once the results for the (bare) $G_u$ and $G_d$ from equations (\ref{eq:GuDelta}) and (\ref{eq:GdDelta}) have been obtained in this way, they can be used in the same anti-resummation rules (\ref{eq:expandG0prop})-(\ref{eq:calpha0exp}) to perform the calculation of any observables starting from bare tadpole-free perturbation theory, including the pole masses of Higgs scalars and the superpartners.

\section{Outlook\label{sec:outlook}}
\setcounter{equation}{0}
\setcounter{figure}{0}
\setcounter{table}{0} 
\setcounter{footnote}{1}

In this paper, I have discussed the effective potential, and its minimization condition, in the context of an efficient method based on first calculating in bare perturbation theory, and then renormalizing the results in the \MSbar scheme 
using parameter redefinitions. Applying this to the calculation of other observables in a tadpole-free scheme requires
expanding around the VEVs that minimize the effective potential. Explicit results were provided for expressing the bare effective potential in general theories at two-loop order. For the three-loop case, certain special cases requiring an enlargement of the list of candidate master integrals for an $\epsilon$-finite basis were discussed, and the relevant expansions of the new master integral ${\bf H}'$ were provided.

In quantum field theories, the use of perturbation theory presents a potential ambiguity
due to the freedom to move some terms from the free part of the Lagrangian to the interaction part. In the standard approach, all terms quadratic in the fields are taken to be part of the free Lagrangian, which is used to construct the propagators. In the case of fields corresponding directly to physical particles, this is usually the most natural way to proceed, as it automatically resums contributions of the same order. 

However, in the case of the Goldstone boson fields, I have argued that for some purposes it is best to take the squared mass terms as two-point interaction vertices instead. Thus they are treated as part of the interaction Lagrangian, leading to the prescription of eq.~(\ref{eq:expandGBprop}), undoing the usual resummation of terms that is often taken for granted to be appropriate. There are two key features of the Goldstone bosons at play here. First, the Goldstone boson fields do not correspond to asymptotic physical states, removing one of the motivations for treating their squared masses as part of the propagator. Second, their squared masses are essentially loop-suppressed quantities, since they vanish in the tree-level approximation.

The treatment of Goldstone bosons using the prescription of eq.~(\ref{eq:expandGBprop}) manifestly avoids negative powers and logarithms of the Goldstone-boson squared masses from the start, and is straightforward to implement in terms of master integrals, as worked out in section \ref{sec:Goldstonerules}. Simple rules for dealing with Goldstone-boson squared masses are given in terms of the master integrals at up to three-loop order in the ancillary file {\tt GoldstoneRules.txt}. The resulting minimization condition for the bare effective potential through three-loop order in the Standard Model is given 
in terms of the quantities $\widetilde\Delta^{(n,k)}$ for $n+k \leq 3$, which are provided in the ancillary file 
{\tt SMGB.txt}. This form of the effective potential minimization condition is an essential ingredient in computations
in the tadpole-free scheme of other quantities including the three-loop pole masses of the Higgs, $W$, and $Z$ bosons. Those results will be reported on elsewhere. 

%%\color{blue}%%NEW
It is less clear whether the anti-resummation prescription can be usefully applied to problems in which the Higgs field deviates far from its minimum, including both the symmetric phase and the opposite limit of very large field strength. In such cases, $G$ may no longer be viewed as a loop-suppressed quantity, and the perturbative expansion in it may no longer be valid. 
Note that the anti-resummed effective potential does not have an imaginary part to encode the information about the Weinberg--Wu \cite{Weinberg:1987vp} instability away from the minimum. 
Furthermore, as noted at the end of section \ref{sec:StandardModel}, in the anti-resummation prescription the renormalized effective potential itself contains infrared poles in $\epsilon$ starting even at 1-loop order. This is not a problem of principle, because the effective potential is not itself a gauge-invariant observable. However, it is a challenge, which may deserve further thought, to see whether the anti-resummation prescription might still be useful in situations involving large deviations from equilibrium.
%%\color{black}%%ENDNEW

Acknowledgments: I thank an anonymous referee for useful comments. This work is supported in part by the National Science Foundation grant with award numbers 2310533 and 2609362.

%%%%%%%%%%%%%%%%%%%%%%%%%%%%%%%%%%%%%%%%%%%%%%%%%%%%%%%%%%%%%%%%%%%%

\end{document}